\documentclass[english]{article}
\usepackage{lmodern}
\renewcommand{\sfdefault}{lmss}

\usepackage[T1]{fontenc}
\usepackage[a4paper]{geometry}
\usepackage{dsfont}
\usepackage{color}
\usepackage{babel}
\usepackage{amsmath}
\usepackage{amsthm}
\usepackage{makecell}
\usepackage{amssymb}
\usepackage{stmaryrd}
\usepackage{graphicx}
\usepackage{setspace}
\usepackage{esint}
\usepackage{caption}
\newcommand*\rot{\rotatebox{90}}
\usepackage[authoryear]{natbib}
\usepackage[unicode=true,bookmarks=true,bookmarksnumbered=false,bookmarksopen=false,breaklinks=false,pdfborder={0 0 0},pdfborderstyle={},backref=page,colorlinks=true]{hyperref}
\hypersetup{pdftitle={Improving Tree Probability Estimation with Stochastic Optimization and Variance Reduction},pdfauthor={Tianyu Xie, Musu Yuan, Minghua Deng, Cheng Zhang},linkcolor=RoyalBlue,citecolor=RoyalBlue}
\usepackage{float}
\usepackage{subfig}
\usepackage[dvipsnames,svgnames,x11names,hyperref]{xcolor}
\usepackage{nicefrac}

\usepackage{epstopdf}
\usepackage{url} 
\usepackage{caption}
\usepackage{multirow}
\usepackage{amssymb}
\usepackage{multicol}
\usepackage{booktabs}
\usepackage{tikz}
\usepackage{bbm}
\usepackage{bm}
\usetikzlibrary{positioning, bayesnet}
\usepackage{algorithm2e}
\RestyleAlgo{ruled}

\usepackage{amsmath,amsfonts,bm}

\def\eqref#1{equation~\ref{#1}}

\def\1{\bm{1}}

\DeclareMathAlphabet{\mathsfit}{\encodingdefault}{\sfdefault}{m}{sl}
\SetMathAlphabet{\mathsfit}{bold}{\encodingdefault}{\sfdefault}{bx}{n}

\newcommand{\E}{\mathbb{E}}

\DeclareMathOperator*{\argmax}{arg\,max}

\newtheorem{theorem}{Theorem}

\newtheorem{definition}{Definition}

\usepackage[textsize=tiny]{todonotes}

\makeatletter
\newcommand{\myfnsymbol}[1]{%
  \expandafter\@myfnsymbol\csname c@#1\endcsname
}
\newcommand{\@myfnsymbol}[1]{%
  \ifcase #1
  \or 1
  \or 2
  \or 3
  \or 4
  \or \TextOrMath{\textasteriskcentered}{*}
  \or \TextOrMath{\textdagger}{\dagger}
  \fi
}
\newcommand{\affiliationA}{\@myfnsymbol{1}}
\newcommand{\affiliationB}{\@myfnsymbol{2}}
\newcommand{\affiliationC}{\@myfnsymbol{3}}
\newcommand{\affiliationD}{\@myfnsymbol{4}}
\newcommand{\equalcontribution}{\@myfnsymbol{5}}
\newcommand{\correspondingA}{\@myfnsymbol{6}}
\makeatother

\begin{document}

\title{Improving Tree Probability Estimation with Stochastic Optimization and Variance Reduction}

\author{
Tianyu Xie\textsuperscript{\affiliationA,\equalcontribution},
Musu Yuan\textsuperscript{\affiliationB,\equalcontribution},
Minghua Deng\textsuperscript{\affiliationC},
Cheng Zhang\textsuperscript{\affiliationD,\correspondingA}
}

\date{
}

\renewcommand{\thefootnote}{\myfnsymbol{footnote}}
\maketitle
\footnotetext[1]{School of Mathematical Sciences, Peking University,
   Beijing, 100871, China. Email: tianyuxie@pku.edu.cn}%
\footnotetext[2]{Center for Quantitative Biology, Peking University,
   Beijing, 100871, China. Email: yuanmusu@pku.edu.cn}%
\footnotetext[3]{Center for Quantitative Biology, School of Mathematical Sciences, and Center for Statistical Science, Peking University,
   Beijing, 100871, China. Email: dengmh@math.pku.edu.cn}%
\footnotetext[4]{School of Mathematical Sciences and Center for Statistical Science, Peking University, Beijing, 100871, China. Email: chengzhang@math.pku.edu.cn}
\footnotetext[5]{These authors contributed equally to this work.}%
\footnotetext[6]{Corresponding author}%

\setcounter{footnote}{0}
\renewcommand{\thefootnote}{\fnsymbol{footnote}}

\begin{abstract}
Probability estimation of tree topologies is one of the fundamental tasks in phylogenetic inference.
The recently proposed subsplit Bayesian networks (SBNs) provide a powerful probabilistic graphical model for tree topology probability estimation by properly leveraging the hierarchical structure of phylogenetic trees.
However, the expectation maximization (EM) method currently used for learning SBN parameters does not scale up to large data sets.
In this paper, we introduce several computationally efficient methods for training SBNs and show that variance reduction could be the key for better performance.
Furthermore, we also introduce the variance reduction technique to improve the optimization of SBN parameters for variational Bayesian phylogenetic inference (VBPI).
Extensive synthetic and real data experiments demonstrate that our methods outperform previous baseline methods on the tasks of tree topology probability estimation as well as Bayesian phylogenetic inference using SBNs.
\end{abstract}




\section{Introduction}
One of the most fundamental goals in modern computational biology is to reconstruct the evolutionary history and relationships among individuals or groups of biological entities.
The reconstructed phylogeny is of great interest to many downstream tasks concerning evolutionary and genomics research.
One commonly used statistical approach for phylogeny reconstruction is Bayesian phylogenetic inference.
Given properly aligned sequence data (e.g., DNA, RNA, Protein, etc.) and a probabilistic evolutionary model that describes the stochastic processes of these heritable traits, Bayesian phylogenetics provides principled ways to quantify the uncertainty of the evolutionary process in terms of the posterior probabilities of phylogenetic trees \citep{Huelsenbeck2001MRBAYESB}.

As a classical Bayesian inference method, Markov chain Monte Carlo (MCMC) is commonly used to draw samples from the phylogenetic posteriors \citep{Yang1997BayesianPI,Mau1999BayesianPI}. 
The posterior probabilities of phylogenetic trees are then typically estimated with simple sample relative frequency (SRF), based on those MCMC samples.
However, SRF does not support trees beyond observed samples (i.e. simply sets the probabilities of unsampled trees to zero) and is prone to unstable estimates for low-probability trees \citep{larget2013estimation}.
As a result, reliable estimation with SRF often requires impractically large sample sizes, especially when the tree space is large.
Recent works \citep{Hhna2012-pm,larget2013estimation} show that harnessing the similarity of tree topologies could be helpful for alleviating these problems.
However, the conditional independence assumption of separated subtrees therein is often too strong to provide accurate approximations for posteriors inferred from real data \citep{whidden2015quantifying}.

Inspired by these previous works, \citet{zhang2018sbn} proposed a general framework for tree topology probability estimation by introducing a novel probabilistic graphical model called subsplit Bayesian networks (SBNs).
Utilizing more sophisticated local topological structures, SBNs relax the conditional clade independence assumption \citep{larget2013estimation} and therefore can provide a rich family of distributions over the entire tree space.
Moreover, these flexible tree space distributions provided by SBNs were later on integrated into variational Bayesian phylogenetic inference (VBPI), which is an alternative approximate Bayesian inference method to MCMC that can deliver competitive phylogenetic posterior estimates in a more timely manner \citep{zhang2018vbpi,zhang2020vbpinf}.

While SBNs have proved effective for tree topology probability estimation, current approaches rely on the celebrated expectation maximization (EM) algorithm to learn SBN parameters, which typically requires expensive full batch computation in each iteration and may get stuck at some local mode due to its monotonical behavior.
In this paper, we propose several advanced techniques for efficient training of SBNs that scale up to large data sets.
Although stochastic expectation maximization (SEM) has been proposed to scale up the EM algorithm \citep{cappe2009sem}, we find that a naive implementation of SEM for SBN training may deteriorate the estimates due to the large variance of stochastic updates.
Fortunately, this issue can be remedied by incorporating the variance reduction technique \citep{chen2018semvr} which leads to a variance reduced stochastic expectation maximization algorithm for SBN training that we call SEMVR.
Although the full batch gradient based method tends to be slower than full batch EM due to the ignorance of SBN structures, we find that the stochastic gradient method, when combined with variance reduction as suggested by \citet{johnson2013svrg}, can provide comparable training efficiency for SBNs and refer to it as SVRG. We show that SEMVR and SVRG evidently outperform the original EM method and other tree topology probability estimation methods on both synthetic data and a benchmark of challenging phylogenetic posterior estimation problems.
We also find variance reduction to be useful for learning the SBN parameters in VBPI with the reweighted wake-sleep (RWS) gradient estimator.
The corresponding variance reduced gradient estimator, which we call RWSVR, can provide more stable gradient estimates for the SBN parameters which eventually improves the approximation accuracy of tree topology posteriors.
Experiments on a benchmark of real data variational Bayesian phylogenetic inference problems demonstrate the advantage of RWSVR over RWS.

The rest of the paper is organized as follows. In section \ref{sec2-background}, we introduce basic concepts and notations of phylogenetic models and subsplit Bayesian networks (SBNs).
In section \ref{sec-methods}, we propose several improved techniques for training SBNs in the tasks of tree topology probability estimation and variational Bayesian phylogenetic inference.
In section \ref{sec4-experiments}, we compare the proposed methods to existing baselines on both synthetic data and real data problems.
We conclude with a discussion in section \ref{sec:conclusion}.
Throughout this paper, we mainly focus on SBN learning on unrooted tree topologies and will use `tree topology' for unrooted tree topology unless otherwise specified.

\section{Background}\label{sec2-background}
\subsection{Phylogenetic Trees} 
Phylogenetic trees are the fundamental structures for describing the evolutionary history of a family of species.
Generally speaking, a phylogenetic tree $T$ is defined as a tree topology $\tau$ and a set of corresponding branch lengths $\boldsymbol{l}$ for the edges on $\tau$. 

The tree topology $\tau$ is a bifurcating tree graph $\left(N(\tau), E(\tau)\right)$, where $N(\tau)$ and $E(\tau)$ are the set of nodes and edges respectively.
Each node in $N(\tau)$ has 1 to 3 neighbors.
Nodes that have 1 neighbor are called \textit{leaf nodes} and the others are called \textit{internal nodes}.
For unrooted tree topologies, all edges are undirected and all internal nodes have 3 degrees; for rooted tree topologies, there is a special internal node of degree 2 called the \textit{root node} (or \textit{root} for simplicity) and the other internal nodes have 3 degrees.
A leaf node represents an existing species (also known as a \textit{taxon}), while an internal node represents an ancestor species that has existed historically.
The edges in a rooted tree are directed and point away from the root, interpreted as the evolution of species originating from the root.
Note that an unrooted tree topology can be converted to a rooted one when a ``virtual root'' is placed on one of its edges. 

For a tree topology $\tau$, each edge $(u,v)\in E(\tau)$ is associated with a branch length $l_{uv}$.
The set of branch lengths is then denoted as $\bm{l}=\{l_{uv}: (u,v)\in E(\tau)\}$.
The branch length $l_{uv}$ quantifies the intensity of the evolutionary changes between node $u$ and node $v$; or more concretely, it is proportional to the expected number of substitutions per site between the two neighboring nodes. 

\subsection{Phylogenetic Posterior}
The goal of phylogenetic analysis is to reconstruct the phylogenetic tree based on the observed data at the leaf nodes.
In Bayesian phylogenetics, this then amounts to properly estimating the phylogenetic posterior which we describe as follows.
Let the matrix $\bm{Y} = \{Y_1,Y_2,\ldots, Y_S\}\in \Omega^{N\times S}$ be the observed sequence data (e.g. DNA, RNA, protein, etc.), where $S$ is the sequence length, $N$ is the number of taxa that correspond to the leaf nodes and $\Omega$ is the set of all characters.
Given a rooted phylogenetic tree $T = (\tau,\bm{l})$, the probability of observing sequence data $\bm{Y}$  is usually defined on top of a continuous-time Markov chain \citep{jukes1969evolution, tavare1986some}, which is known as the substitution model in the literature.
Let $Q, \eta$ be the transition rate matrix and the stationary distribution of the continuous time Markov chain.
Let $a_{v}^s$ be the state of node $v$ at site $s$, the transition probability along a branch $(u,v)$ at site $s$ given by the substitution model is $P_{a^s_u a^s_v}(l_{uv}) = \left(e^{l_{uv}Q}\right)_{a^s_u, a^s_v}$.
Using the Markov property, the probability of each site observation $Y_s$ is then defined as the probability distribution of the leaf nodes by marginalizing out all possible states of the unobserved internal nodes as follows
\begin{equation}\label{one-site-prob}
p(Y_s|\tau,\bm{l}) = \sum_{a^s}\eta(a^s_\rho)\prod_{(u,v)\in E(\tau)} P_{a^s_u a^s_v}(l_{uv}),
\end{equation}
where $\rho$ is the root node and $a^s$ ranges all extensions of $Y_s$ to the internal nodes with $a^s_u$ being the assigned character of node $u$ ($a^s_u = Y_{u,s}$ if $u$ is a leaf node, where $Y_{u,s}$ is the observed character of node $u$ at site $s$). For an unrooted tree topology $\tau$, (\ref{one-site-prob}) also provides a valid probability when placing the root $\rho$ on an arbitrary edge. In fact, the probability (\ref{one-site-prob}) is irrelevant to the position of the root if the continuous-time Markov model is time reversible \citep{felsenstein1981evolutionary}, which is known as the pulley principle.

Assuming that all the $S$ sites are identically distributed and independently evolved, the likelihood function can be expressed as
\begin{equation}\label{full-prob}
p(\bm{Y}|\tau,\bm{l}) =\prod_{s=1}^S p(Y_s|\tau,\bm{l})  = \prod_{s=1}^S  \sum_{a^s}\eta(a^s_\rho)\prod_{(u,v)\in E(\tau)} P_{a^s_u a^s_v}(l_{uv}).
\end{equation}
The phylogenetic likelihood function defined in (\ref{full-prob}) can be efficiently evaluated using Felsenstein's pruning algorithm \citep{felsenstein2004inferring}.
Given an appropriate prior distribution $p(\tau,\bm{l})$ on the phylogenetic tree, the phylogenetic posterior is
\begin{equation}\label{posterior}
p(\tau,\bm{l}|\bm{Y}) = \frac{p(\boldsymbol{Y}|\tau, \boldsymbol{l})p(\tau,\bm{l})}{p(\bm{Y})}\propto p(\boldsymbol{Y}|\tau, \boldsymbol{l})p(\tau,\bm{l}).
\end{equation}

\subsection{Subsplit Bayesian Networks}\label{background-sbn}
Subsplit Bayesian networks, as proposed by \citet{zhang2018sbn} recently, is an expressive graphical model that provides a flexible family of distributions over tree topologies.
Let $\mathcal{X}$ be the set of $N$ labeled leaf nodes. We call a nonempty set $C$ of $\mathcal{X}$ a \textit{clade}.
The set of all clades of $\mathcal{X}$, i.e. $\mathcal{C}(\mathcal{X}) = \{C|C\subset \mathcal{X}, C\neq \emptyset\}$, then forms a totally ordered set with a total order $\succ$ (e.g. lexicographical order) defined on it.
An ordered pair of clades $(W,Z)$ is called a subsplit of a clade $C$ if it is a bipartition of $C$, i.e. $W\succ Z, W\cap Z=\emptyset$ and $W\cup Z=C$.
A subsplit Bayesian network on $\mathcal{X}$ is then defined as follows.
\begin{definition}[Subsplit Bayesian Network]
A subsplit Bayesian network (SBN) $\mathcal{B}_{\mathcal{X}}$ on a leaf node set $\mathcal{X}$ of size $N$ is defined as a Bayesian network whose nodes take on subsplit or singleton clade values of $\mathcal{X}$, and has the following properties: (a) The root node of $\mathcal{B}_{\mathcal{X}}$ takes on subsplits of the entire labeled leaf node set $\mathcal{X}$; (b) $\mathcal{B}_{\mathcal{X}}$ contains a full and complete binary tree network $B^\ast_{\mathcal{X}}$ as a subnetwork; (c) The depth of $B_{\mathcal{X}}$ is $N-1$, with the root counted as depth $1$.
\end{definition}

\begin{figure}
\begin{center}
\input{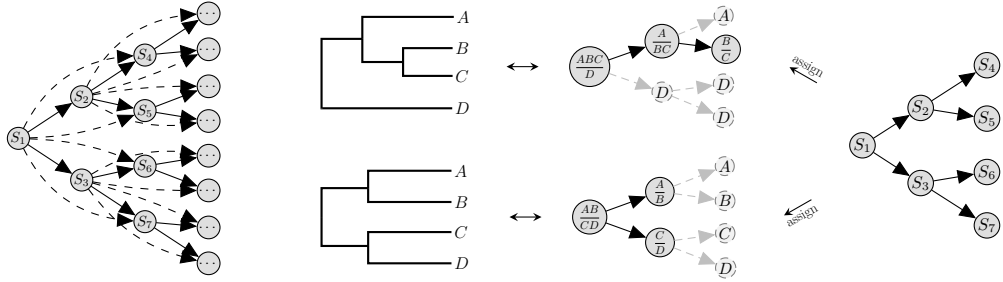}
\caption{Subsplit Bayesian networks and a simple example for a leaf set of 4 taxa (denoted by $A,B,C,D$ respectively).
{\bf Left:} General subsplit Bayesian networks. The solid full and complete binary tree network is $B^\ast_{\mathcal{X}}$.
The dashed arrows represent the additional dependence for more expressiveness.
{\bf Middle Left:} Examples of (rooted) phylogenetic trees that are hypothesized to model the evolutionary history of the taxa.
{\bf Middle Right:} The corresponding subsplit assignments for the trees.
For ease of illustration, subsplit $(Y,Z)$ is represented as $\frac{Y}{Z}$ in the graph.
{\bf Right:} The SBN for this example, which is $\mathcal{B}_\mathcal{X}^\ast$ in this case.
This Figure is from \citet{zhang2018sbn}.
}\label{fig:sbn}
\end{center}
\end{figure}

The unique $B^\ast_{\mathcal{X}}$ contained in all SBNs on $\mathcal{X}$ provides a universal indexing for the nodes in all SBNs on $\mathcal{X}$. 
This is achieved by denoting the root node with $S_1$ and the two children of $S_i$ with $S_{2i}$ (top) and $S_{2i+1}$ (bottom) recursively, for any internal node $S_i$ (see the left plot in Figure \ref{fig:sbn}). 
To illustrate how SBNs provide distributions over tree topologies, we need the definition of compatibility. 

\begin{definition}[Compatible Subsplit Assigment]
A subsplit $(W,Z)$ is said to be compatible with a clade $C$ if $W\cup Z=C$.
A full subsplit assignment $\{S_i=s_i\}_{i\geq 1}$ is compatible if for any interior node assignment $s_i = (W_i,Z_i)$, the child node assignments $s_{2i}, s_{2i+1}$ are compatible with $W_i, Z_i$ respectively, for any non-singleton clade $W_i$ or $Z_i$.
\end{definition}

Let $\mathcal{T}_{\mathcal{X}}$ be the set of rooted tree topologies with leaf labels in $\mathcal{X}$.
According to Lemma 1 in \citet{zhang2018sbn}, there exists a bijection between rooted tree topologies $\tau \in \mathcal{T}_{\mathcal{X}}$ and compatible subsplit assignments of the nodes of $\mathcal{B}_{\mathcal{X}}$. 
Furthermore, the corresponding compatible subsplit assignments $\{S_i=s_i\}_{i\geq 1}$ for a rooted tree topology $\tau \in \mathcal{T}_{\mathcal{X}}$ can be obtained by following the splitting process of $\tau$ and assigning the subsplits to the corresponding nodes along the way from the root node to the leaf nodes (see the middle plots in Figure \ref{fig:sbn}). 
The SBN-induced probability of $\tau$, therefore, takes the following form
\begin{equation}\label{sbn-prob-rooted}
q(\tau) = p(S_1=s_1)\prod_{i>1}p(S_i=s_i|S_{\pi_i} = s_{\pi_i}),
\end{equation}
where $\pi_i$ denotes the set of indices of the parent nodes of $S_i$.
As Bayesian networks, \eqref{sbn-prob-rooted} defines proper distributions over $\mathcal{T}_{\mathcal{X}}$ as long as the conditional probabilities are consistent, which is a common property of Bayesian networks.

The SBN framework also generalizes to unrooted tree topologies, which are the most common type of tree topologies in phylogenetics.
Let $\mathcal{T}_{\mathcal{X}}^{\mathrm{u}}$ be the set of unrooted tree topologies with leaf labels in $\mathcal{X}$.
For an unrooted tree topology $\tau^{\mathrm{u}} \in \mathcal{T}_{\mathcal{X}}^{\mathrm{u}}$, let $\mathcal{R}(\tau^{\mathrm{u}}) = \{\tau_{e}:e\in E(\tau^{\mathrm{u}})\}$ be the set of rooted trees that are constructed by placing a ``virtual root'' on one edge of $\tau^{\mathrm{u}}$. Then the SBN-induced probability of $\tau^{\mathrm{u}}$ is given by
\begin{equation}\label{sbn-prob-unrooted}
q(\tau^{\mathrm{u}}) = \sum_{\tau\in \mathcal{R}(\tau^{\mathrm{u}})} q(\tau) = \sum_{e\in E(\tau^{\mathrm{u}})}q(\tau_{e}).
\end{equation}
The above \eqref{sbn-prob-unrooted} can be viewed as a marginal probability where the ``missing" root node of $\tau^{\mathrm{u}}$ is integrated over all possible positions (i.e., the edges).
As $\{\mathcal{R}(\tau^{\mathrm{u}}): \tau^{\mathrm{u}}\in \mathcal{T}^{\mathrm{u}}_{\mathcal{X}}\}$ naturally defines a partition over $\mathcal{T}_{\mathcal{X}}$, \eqref{sbn-prob-unrooted} also provides proper distributions over $\mathcal{T}^{\mathrm{u}}_{\mathcal{X}}$. 

In practice, SBNs are often parameterized according to the \textit{conditional probability sharing} principle where the conditional probability distributions (CPDs) for parent-child subsplit pairs are shared across the SBNs, regardless of their locations \citep{zhang2018sbn}.
More specifically, denote the set of observed root subsplits as $\mathbb{S}_{\mathrm{r}}$ and the set of observed parent-child subsplit pairs as $\mathbb{S}_{\mathrm{ch|pa}}$. The parameters of SBNs (i.e., the support of CPDs) are then $c=\{c_{s}: s\in \mathbb{S}_{\mathrm{r}}\}\cup \{c_{s|t}: s|t\in \mathbb{S}_{\mathrm{ch|pa}}\}$ where
\begin{equation}\label{eq:sbn_param}
p(S_1=s_1) = c_{s_1},\quad p(S_i=s|S_{\pi_i}=t) = c_{s|t}, \; \forall i>1. 
\end{equation}

\section{Improved Techniques for Training SBNs}\label{sec-methods}
While SBNs provides a rich family of distributions for tree topology probability estimation, learning SBN parameters is currently done via the celebrated expectation maximization (EM) algorithm that requires expensive full batch computation in each iteration, making it challenging to scale up to large data sets.
Moreover, EM is also prone to get stuck at local modes due to monotonicity.
In this section, we propose several computationally efficient methods to alleviate these issues.
We first introduce a stochastic EM algorithm for training SBNs, together with a variance reduction technique for more stable gradient estimates.
We then introduce a variance reduced stochastic gradient ascent method that can provide comparable performance although its full batch version tends to be slower than EM.
Finally, we show that variance reduction is useful for variational Bayesian phylogenetic inference and introduce an improved RWSVR gradient estimator based on it.

\subsection{Tree Topology Probability Estimation}\label{sec-sbn}
Given a sample of unique unrooted tree topologies $\mathcal{D} = \{\tau_k\}_{k=1}^K$ with corresponding weights $\mathcal{W}=\{w_k\}_{k=1}^{K}$ (e.g., SRF of sampled trees from a phylogenetic MCMC run), the tree topology probability estimation task requires accurate approximation for tree probabilities over the entire tree space (e.g., accurate posterior estimates for all trees, including those that are unsampled).
SBNs prove effective for this task by providing a rich family of distributions over the tree topology space that generalize beyond the sampled trees.
Treating the missing root node as a latent variable, SBNs are currently learned via the expectation maximization (EM) algorithm \citep{zhang2018sbn}.

\subsubsection{The EM Algorithm}
As discussed in Section \ref{background-sbn}, SBNs for unrooted tree topologies is indeed a latent variable model with the root node being unobserved. 
Learning SBNs for tree topology probability estimation, therefore can be achieved via the celebrated EM algorithm as follows.
Let $\tau_{k,e}$ denote the corresponding rooted tree when the root is placed at edge $e\in E(\tau_k)$ and $\{S_i = s^{e}_{i,k}\}_{i\geq 1}$ denote the subsplit assignment for $\tau_{k,e}$. 
Given $\mathcal{D}$ with weights $\mathcal{W}$, the data log-likelihood of the CPDs $c$ is
\begin{equation}\label{eq:full-data-loglikelihood}
\mathcal{L}(c;\mathcal{D},\mathcal{W}) =\sum_{k=1}^K w_k\log p(\tau_k|c)=\sum_{k=1}^K w_k\log\left(\sum_{e\in E(\tau_k)}p(\tau_{k,e}|c)\right).
\end{equation}

In the E-step, given the current estimated CPDs $\hat{c}$, we compute the full-sample $Q$-function which is a lower bound of the data log-likelihood function $\mathcal{L}$ that can be easily maximized. More specifically, for each unrooted tree topology $\tau_k$, we compute the single-sample $Q$-function, i.e., the expected complete data log-likelihood of CPDs $c$ 
\[
Q_k(c;\hat{c}) := \mathbb{E}_{e|\tau_k, \hat{c}}\log p(\tau_{k,e}|c) = \sum_{e\in E(\tau_k)}p(s_{1,k}^e|\tau_k, \hat{c})\log p(\tau_{k,e}|c),
\]
where $p(s_{1,k}^e|\tau_k, \hat{c})$ is the conditional probability of missing root at $e$.
Based on (\ref{sbn-prob-rooted}) and the parameterization (\ref{eq:sbn_param}), the SBN-induced tree topology log-probability of $\tau_{k,e}$ is
\begin{align*}
\log p(\tau_{k, e}|c) &= \log p(S_1=s_{1,k}^e) + \sum_{i>1}\log p(S_i=s_{i,k}^e|S_{\pi_i}=s_{\pi_i,k}^e)\\
&=\sum_{s\in\mathbb{S}_{\mathrm{r}}}\mathbb{I}(s=s_{1,k}^e)\log c_s + \sum_{i>1}\sum_{s|t\in \mathbb{S}_{\mathrm{ch|pa}}}\mathbb{I}(s=s_{i,k}^e, t=s_{\pi_i, k}^e)\log c_{s|t},
\end{align*}
where $\mathbb{I}$ is the indicator function. If we define the single-sample expected frequency counts (EFCs) of $\tau_k$ by
\begin{equation}\label{m-function}
m_{k, s}(\hat{c}) = \sum_{e\in E(\tau_k)} p(s_{1,k}^e|\tau_k, \hat{c})\mathbb{I}(s=s_{1,k}^e),\ m_{k, s|t}(\hat{c}) = \sum_{e\in E(\tau_k)} p(s_{1,k}^e|\tau_k, \hat{c}) \sum_{i>1}\mathbb{I}(s=s_{i,k}^e, t=s_{\pi_i, k}^e),
\end{equation}
the single-sample $Q$-function takes an explicit form
\begin{equation}\label{q-function}
Q_k(c;\hat{c}) = \sum_{e\in E(\tau_k)}p(s_{1,k}^e|\tau_k, \hat{c})\log p(\tau_{k,e}|c) = \sum_{s\in\mathbb{S}_{\mathrm{r}}}m_{k, s}(\hat{c}) \log c_s + \sum_{s|t\in \mathbb{S}_{\mathrm{ch|pa}}} m_{k, s|t}(\hat{c}) \log c_{s|t}.
\end{equation}
The full-sample $Q$-function then is computed as a weighted sum of its single-sample version
\begin{equation}\label{eq:full-q-function}
Q(c;\hat{c}) = \sum_{k=1}^K w_k Q_k(c;\hat{c}) = \sum_{s\in\mathbb{S}_{\mathrm{r}}}M_{ s}(\hat{c}) \log c_s + \sum_{s|t\in \mathbb{S}_{\mathrm{ch|pa}}} M_{s|t}(\hat{c}) \log c_{s|t},
\end{equation}
where
\begin{equation}\label{eq:sufficient_stats}
M_s(\hat{c}) = \sum_{k=1}^K w_k m_{k,s}(\hat{c}),\quad M_{s|t}(\hat{c}) = \sum_{k=1}^K w_k m_{k,s|t}(\hat{c})
\end{equation}
are the sufficient statistics which we denote by $M(\hat{c})=\{M_s(\hat{c}): s\in \mathbb{S}_{\mathrm{r}}\}\cup \{M_{s|t}(\hat{c}): s|t\in\mathbb{S}_{\mathrm{ch}|\mathrm{pa}}\}$.

In the M-step, we maximize $Q(c;\hat{c})$ to update the estimates of CPDs $c$ which has a simple closed-form solution $c^\ast = \Phi(M(\hat{c})) = \argmax_{c}Q(c,\hat{c})$ as follows
\begin{equation}\label{eq:mstep}
     c_{s}^\ast = \frac{M_{s}(\hat{c})}{\sum_{s'\in\mathbb{S}_\mathrm{r}}M_{s'}(\hat{c})}, \;\forall\; s\in \mathbb{S}_\mathrm{r}, \quad c_{s|t}^{\ast} = \frac{M_{s|t}(\hat{c})}{\sum_{s'|t\in\mathbb{S}_\mathrm{ch|pa}}M_{s'|t}(\hat{c})}, \; \forall\; s|t\in\mathbb{S}_\mathrm{ch|pa}. 
\end{equation}
Let $\hat{c}^{(n)}$ denote the estimates of CPDs at the $n$-th step, the EM algorithm then has the following updating scheme:
\begin{itemize}
    \item E-step: $\forall 1\leq k\leq K, e\in E(\tau_k)$, compute $p(s_{1,k}^e|\tau_k, \hat{c}^{(n)}) = \frac{p(s_{1,k}^e,\tau_k| \hat{c}^{(n)})}{\sum_{e\in E(\tau_k)}p(s_{1,k}^e,\tau_k| \hat{c}^{(n)})}$.
    \item M-step: update the estimates of CPDs as $\hat{c}^{(n+1)} = \Phi(M(\hat{c}^{(n)}))$.
\end{itemize}
When the data are insufficient, we can also incorporate regularization by assuming Dirichlet prior on CPDs and we call this algorithm EM-$\alpha$ (see more details in Appendix \ref{app-emalpha}).

While effective, the E-step in EM requires full batch computation which would become costly when the number of tree topologies $K$ is large.
Moreover, the parameter estimates obtained by EM are prone to get stuck in a stationary point rather than a global or local maximizer and may suffer from slow convergence especially when the information matrix vanishes \citep{mclachlan2007em}. These phenomena could be quite remarkable for SBN training due to the highly non-convex SBN-based probabilities of tree topologies.
To accelerate computation and ease optimization, we propose two stochastic algorithms, together with their variance-reduced variants, for tree topology probability estimation via SBNs.

\subsubsection{Stochastic EM and Variance Reduced Stochastic EM}
One approach to scale up EM to large data sets is stochastic EM \citep{cappe2009sem, cappe2011online}.
Assuming the complete data likelihood belongs to an exponential family, stochastic EM (SEM) replaces the expensive full batch expectation in the E-step by a stochastic approximation that sequentially updates the vector of sufficient statistics while keeping the M-step unchanged.
Let $m_k(\hat{c})=\{m_{k,s}(\hat{c}): s\in \mathbb{S}_{\mathrm{r}}\}\cup \{m_{k,s|t}(\hat{c}): s|t\in\mathbb{S}_{\mathrm{ch}|\mathrm{pa}}\}$ be the EFCs for $\tau_k$ as defined in \eqref{m-function}.
Following \citet{cappe2009sem}, we can modify the E-step by maintaining an exponential moving average of EFCs as follows
\begin{equation}\label{sem}
\Bar{M}^{(n+1)} = (1-\rho_{n+1})\Bar{M}^{(n)}+\rho_{n+1} m_{\mathcal{B}}(\hat{c}^{(n)}),
\end{equation}
where $\bar{M}^{(n)}$ is the estimated sufficient statistics at iteration $n$, $m_{\mathcal{B}}(\hat{c}) = \sum_{b=1}^B m_{k_b^{(n+1)}}(\hat{c})/B$ is the mean EFCs from a mini-batch $\mathcal{B} = \{\tau_{k_b^{(n+1)}}\}_{b=1}^B$ that is sampled from $\mathcal{D}$ according to the corresponding weights $\mathcal{W}$, and $\{\rho_{n}\}$ is a decreasing sequence of positive learning rates.

In practice, the variance introduced by the mini-batch approximation can be high, leading to a slow asymptotic convergence rate of $O(1/\sqrt{N})$ for SEM \citep{cappe2009sem}, where $N$ is the number of iterations.
Inspired by variance reduction techniques for stochastic gradient descent methods \citep{SAG,SAGA,johnson2013svrg}, \citet{chen2018semvr} proposed a variance reduction strategy for stochastic EM (SEMVR) that uses
infrequently computed batch expectations as control variates.
More specifically, SEMVR runs $T$ mini-batch iterations in each epoch, with iteration $t$ in epoch $h$ indexed as $(h,t)$.
At the beginning of epoch $h$, a full-batch computation is done to get the sufficient statistics $M(\hat{c}^{(h,0)})$, where $\hat{c}^{(h,0)}$ are the estimates of CPDs at the end of the previous epoch.
Let $\bar{M}^{(h,t)}$ be the estimated sufficient statistics at iteration $(h,t)$.
The variance reduced E-step is then given by
\begin{equation}\label{e-step}
\Bar{M}^{(h,t+1)}=(1-\rho)\Bar{M}^{(h,t)}+\rho\left(m_{\mathcal{B}}(\hat{c}^{(h,t)}) - m_{\mathcal{B}}(\hat{c}^{(h,0)})+M(\hat{c}^{(h,0)})\right),
\end{equation}
where $m_{\mathcal{B}}(\hat{c})$ is the mini-batch EFCs as in \eqref{sem} and $\rho$ is a constant learning rate. To ensure the existence of solutions in the M-step, we perform a clipping operation
\begin{equation}\label{clip-step}
\Bar{M}^{(h,t+1)}_{\ast} = \max\left\{\Bar{M}^{(h,t+1)},\lambda\right\}
\end{equation}
where the clipping threshold $\lambda$ is a small positive number and $\max$ refers to elementwise maximization. We call $\Bar{M}^{(h,t)}_{\ast}$ the clipped average EFCs at iteration $(h,t)$.

The M-steps for SEM and SEMVR are similar to the M-step for the standard full batch EM as defined in \eqref{eq:mstep}, where the full-batch sufficient statistics $M$ is replaced by the average EFCs $\bar{M}$ obtained in \eqref{sem} and the clipped average EFCs $\bar{M}_\ast$ in \eqref{clip-step} respectively. 
We summarize the SEMVR approach in Algorithm \ref{alg-semvr}.
For SEM and SEMVR, a Dirichlet prior can also be added on CPDs $c$ as regularization as in EM, and we call the regularized versions SEM-$\alpha$ and SEMVR-$\alpha$ respectively.

\begin{algorithm}[t]
\caption{The SEMVR Algorithm for SBN-based Tree Topology Probability Estimation}\label{alg-semvr}
\KwIn{Tree topology sample $\{\tau_k\}_{k=1}^K$ with weights $\{w_k\}_{k=1}^K$; learning rate $\rho$; minibatch size $B$; number of iterations per epoch $T$; a clipping threshold $\lambda$.}
Initialize $\hat{c}^{(0,0)}$; let $\Bar{M}^{(0,0)} = M(\hat{c}^{(0,0)})$, $h=0$\;
\While{not converge}{
Calculate the full-sample EFCs $M(\hat{c}^{(h,0)})$\;
\For{$t=0,\ldots,T-1$}{
Sample $B$ tree topologies $\mathcal{B}=\{\tau_{k_b}\}_{b=1}^B$ from $\{\tau_k\}_{k=1}^K$ (with replacement) according to their weights $\{w_k\}_{k=1}^K$\;
\textbf{E-step.} Compute $\Bar{M}^{(h,t+1)} = (1-\rho)\Bar{M}^{(h,t)} + \rho\left(m_{\mathcal{B}}(\hat{c}^{(h,t)})-m_{\mathcal{B}}(\hat{c}^{(h,0)})+ M(\hat{c}^{(h,0)})\right)\}$\;
\textbf{Clipping.} Compute $\Bar{M}^{(h,t+1)}_{\ast} = \max\left\{\Bar{M}^{(h,t+1)},\lambda\right\}$\;
\textbf{M-step.} Update $\hat{c}^{(h, t+1)} = \Phi(\Bar{M}^{(h,t+1)}_\ast)$ (defined in \eqref{eq:mstep})\;
}
$\Bar{M}^{(h+1,0)} \leftarrow \Bar{M}^{(h,T)}$; $\hat{c}^{(h+1, 0)}\leftarrow \hat{c}^{(h, T)}$;\ $h\leftarrow h+1$\;
}
\end{algorithm}

\subsubsection{Stochastic Variance Reduced Gradient}
It's also possible to scale up learning of SBNs via gradient-based methods.
To do that, we first accommodate the simplex constraints of CPDs with the following reparameterization
\begin{equation}\label{sbn-param}
c_{s} = \frac{\exp(\phi_{s})}{\sum_{s'\in \mathbb{S}_{\mathrm{r}}}\exp(\phi_{s'})},\ s\in \mathbb{S}_{\mathrm{r}};\quad c_{s|t} = \frac{\exp(\phi_{s|t})}{\sum_{s': s'|t\in \mathbb{S}_{\mathrm{ch|pa}}}\exp(\phi_{s'|t})},\ s|t\in \mathbb{S}_{\mathrm{ch|pa}}.
\end{equation}
We call $\bm{\phi} = \{\phi_s: s\in \mathbb{S}_{\mathrm{r}}\cup \mathbb{S}_{\mathrm{ch|pa}}\}$ the \textit{latent parameters} of CPDs.
The log-likelihood function of $\bm{\phi}$, therefore, takes the following form
\begin{equation}\label{emp-likelihood}
\mathcal{L}(\bm{\phi};\mathcal{D},\mathcal{W}) = \sum_{i=1}^K w_i \log q_{\bm{\phi}}(\tau_i).
\end{equation}

To learn SBNs, we can maximize the log-likelihood function (\ref{emp-likelihood}) via stochastic gradient ascent (SGA).
Let $\bm{\phi}^{(h,t)}$ be the estimates of $\bm{\phi}$ at iteration $t$ in epoch $h$, 
the SGA algorithm updates the parameters with the following stochastic gradient 
\begin{equation}\label{sg}
\hat{G}_{\mathcal{B}}(\phi^{(h,t)}) = \frac{1}{B}\sum_{b=1}^B \nabla_{\bm{\phi}}\log q_{\bm{\phi}}(\tau_{k_b})\big|_{\bm{\phi}=\bm{\phi}^{(h,t)}},
\end{equation}
where a minibatch $\mathcal{B} = \{\tau_{k_b}\}_{b=1}^B$ is sampled from $\mathcal{D}$ according to their weights $\mathcal{W}$.
The efficiency of SGA can be further improved when combined with variance reduction techniques as mentioned in the previous section.
For example, the stochastic variance reduced gradient (SVRG) algorithm uses infrequently computed full-batch gradient as control variates for variance reduction \citep{johnson2013svrg}.
At the beginning of epoch $h$, a full-batch gradient $\nabla_{\bm{\phi}}\mathcal{L}(\bm{\phi}^{(h,0)};\mathcal{D},\mathcal{W})$ is computed, where $\bm{\phi}^{(h,0)}$ is the estimates of latent parameters at the end of the previous epoch.
At iteration $t$ in epoch $h$, the parameters are updated with variance reduced stochastic gradient 
\begin{equation}\label{svrg}
\hat{G}(\bm{\phi}^{(h,t)}) = \hat{G}_{\mathcal{B}}(\bm{\phi}^{(h,t)}) - \hat{G}_{\mathcal{B}}(\bm{\phi}^{(h,0)}) + \nabla_{\bm{\phi}}\mathcal{L}(\bm{\phi}^{(h,0)};\mathcal{D},\mathcal{W}).
\end{equation}
In practice, we may choose a constant step size for SVRG, since the gradient estimate in \eqref{svrg} approaches zero as the algorithm converges.

\begin{table}[t]
\centering
\caption{The numbers of likelihood evaluations and parameter updates per epoch and comparisons for different methods. We assume the batch size for stochastic optimization is $B$, the number of iterations per epoch is $T$, and the total sample size of the training set is $K$.
``-'' means that there is no need for variance reduction because EM has no variance.
}
\label{tab:computation-complexity}
\begin{tabular}{cccccc}
\toprule
Method&EM&SEM&SEMVR&SGA&SVRG\\
\midrule
\makecell{\# likelihood computations}&$K$&$TB$&$K+TB$&$TB$&$K+TB$\\
\makecell{\# parameter updates}&$1$&$T$&$T$&$T$&$T$\\
mini-batch & $\times$&\checkmark&\checkmark&\checkmark&\checkmark\\
variance reduction&-&$\times$&\checkmark&$\times$&\checkmark\\
\bottomrule
\end{tabular}
\end{table}

\subsection{Variational Bayesian Phylogenetic Inference}
The flexible tree topology distributions provided by SBNs have also been used in
variational Bayesian phylogenetic inference (VBPI), which is a recent variational approach for approximate phylogenetic posterior estimation \citep{zhang2018vbpi, zhang2020vbpinf, zhang2022variational}.
Unlike tree topology probability estimation, in VBPI we do not have samples of trees in advance.
The supports of CPDs in SBNs, therefore, are usually estimated using heuristic methods such as ultrafast bootstrap approximation \citep{minh2013ultrafast}.
Combining the expressive SBN-based tree topology distributions $q_{\bm{\phi}}(\tau)$ with a continuous distribution $q_{\bm{\psi}}(\bm{l}|\tau)$ over the branch lengths forms the variational family for phylogenetic trees in VBPI.
The variational approximations are then trained by minimizing the KL divergence using stochastic gradient ascent via efficient Monte Carlo gradient estimators \citep{mnih2016vimco,RWS,VAE}.
While recent progresses on VBPI mainly focus on constructing more flexible distribution families for branch lengths \citep{zhang2020vbpinf}, learning the tree topology parameters using Monte Carlo gradient estimators (e.g., VIMCO and RWS) remains challenging, especially when the variance is large.
To facilitate the learning of SBNs in VBPI, in what follows, we propose a variance reduced reweighted wake-sleep estimator (RWSVR) to stabilize gradient estimation w.r.t the tree topology parameters.  

\subsubsection{The Reweighted Wake-sleep Estimator}
Given a tree topology $\tau$, the conditional branch lengths distribution is often taken to be a diagonal log-normal distribution
\begin{equation}
q_{\bm{\psi}}(\bm{l}|\tau) = \prod_{e\in E(\tau)}p^{\mathrm{Lognormal}}(l_e|\mu_{\bm{\psi}}(\tau, e), \sigma_{\bm{\psi}}(\tau, e)),
\end{equation}
where the mean $\mu_{\bm{\psi}}(\tau, e)$ and the standard deviation $\sigma_{\bm{\psi}}(\tau, e)$ are amortized over the tree topology space based on shared local topological structures (see \citet{zhang2018vbpi} for more details).
The variational family of phylogenetic trees then takes the form $q_{\bm{\phi}, \bm{\psi}}(\tau,\bm{l}) = q_{\bm{\phi}}(\tau)q_{\bm{\psi}}(\bm{l}|\tau)$.

The reweighted wake-sleep estimator is derived when minimizing the inclusive KL divergence from variational approximation to the target posterior,
i.e., $\E_{p(\tau, \bm{l}|\bm{Y})}\log\left(\frac{p(\tau, \bm{l}|\bm{Y})}{q_{\bm{\phi}, \bm{\psi}}(\tau,\bm{l})}\right)$,
which is equivalent to maximizing the likelihood of variational approximation
\begin{equation}
\bm{\phi}^\ast, \bm{\psi}^\ast = \max_{\bm{\phi}, \bm{\psi}} L(\bm{\phi}, \bm{\psi}),\ L(\bm{\phi}, \bm{\psi}) = \E_{p(\tau, \bm{l}|\bm{Y})}\log\left(q_{\bm{\phi}, \bm{\psi}}(\tau,\bm{l})\right).
\end{equation}
Compared with the standard exclusive KL divergence, i.e., $\E_{q_{\bm{\phi}, \bm{\psi}}(\tau,\bm{l})}\log\left(\frac{q_{\bm{\phi}, \bm{\psi}}(\tau,\bm{l})}{p(\tau, \bm{l}|\bm{Y})}\right)$ minimizing the inclusive KL divergence tends to provide better probability estimates for high posterior trees but results in a more challenging optimization problem \citep{MSC}.
The gradients of $L(\bm{\phi}, \bm{\psi})$ w.r.t. the tree topology parameters $\bm{\phi}$ are
\begin{equation}\label{grad}
G(\bm{\phi}) = \E_{p(\tau, \bm{l}|\bm{Y})}\nabla_{\phi}\log q_{\bm{\phi}}(\tau) = \E_{q_{\bm{\phi}, \bm{\psi}}(\tau,\bm{l})}\frac{p(\tau, \bm{l},\bm{Y})/p(\bm{Y})}{q_{\bm{\phi}, \bm{\psi}}(\tau,\bm{l})}\nabla_{\bm{\phi}}\log q_{\bm{\phi}}(\tau),
\end{equation}
where $q_{\bm{\phi}, \bm{\psi}}(\tau,\bm{l})$ is used as the importance distribution.
As the normalizing constant $p(\bm{Y})$ is unknown, we use self-normalized importance sampling (SNIS) instead.
Let the parameter estimates at iteration $t$ in epoch $h$ be $\bm{\phi}^{(h,t)}, \bm{\psi}^{(h,t)}$. Given a sample $\{(\tau^i, \bm{l}^i)\}_{i=1}^R\overset{\mathrm{i.i.d.}}{\sim} q_{\bm{\phi}^{(h,t)}, \bm{\psi}^{(h,t)}}(\tau,\bm{l})$ with unnormalized weights $w^i(\bm{\phi}^{(h,t)}, \bm{\psi}^{(h,t)}) = \frac{p(\tau^i, \bm{l}^i, \bm{Y})}{q_{\bm{\phi}^{(h,t)}, \bm{\psi}^{(h,t)}}(\tau^i,\bm{l}^i)}$, define the SNIS estimator as 
\begin{equation}\label{eq:rws_me}
\hat{G}^{(h,t)}_{R}(\tilde{\bm{\phi}}) = \sum_{i=1}^R \tilde{w}^i(\bm{\phi}^{(h,t)}, \bm{\psi}^{(h,t)}) \nabla_{\bm{\phi}}\log q_{\bm{\phi}}(\tau^i)\big|_{\bm{\phi}=\tilde{\bm{\phi}}},
\end{equation}
where $\tilde{w}^i(\bm{\phi}, \bm{\psi}) = \frac{w^i(\bm{\phi}, \bm{\psi})}{\sum_{j=1}^R w^j(\bm{\phi}, \bm{\psi})}$ is the self-normalized importance weight.
The reweighted wake-sleep (RWS) estimator for $G(\bm{\phi}^{(h,t)})$ is
\begin{equation}\label{rws-estimator}
 \hat{G}_{\mathrm{rws}}(\bm{\phi}^{(h,t)}) = \hat{G}^{(h,t)}_R(\bm{\phi}^{(h,t)}).
\end{equation}

\subsubsection{The Variance Reduced Reweighted Wake-sleep Estimator}
Although SNIS can be helpful for dealing with the unknown normalizing constant, the Monte Carlo estimator used in \eqref{eq:rws_me} may still have high variance, especially when the sample size $R$ is small.
Similarly as in Section \ref{sec-sbn}, we propose to use $\hat{G}_R^{(h,t)}(\bm{\phi}^{(h,0)})$ as control variates for variance reduction, where $\bm{\phi}^{(h,0)}$ is the parameter estimates at the beginning of epoch $h$. Intuitively, as the iteration sequence converges, $\bm{\phi}^{(h,0)}$ will be close to $\bm{\phi}^{(h,t)}$, making $\hat{G}_R^{(h,t)}(\bm{\phi}^{(h,t)})$ and $\hat{G}_R^{(h,t)}(\bm{\phi}^{(h,0)})$ highly correlated.
However, unlike the previous batch setting \citep{johnson2013svrg, chen2018semvr} where an analytical form of the expectation of the gradient estimator is available, we now have to rely on Monte Carlo estimates for $\E_{h,t}\hat{G}_R^{(h,t)}(\bm{\phi}^{(h,0)})$ as it is intractable.

Note that the asymptotic unbiasedness of SNIS estimator implies i) $\{\E_{h,t}\hat{G}^{(h,t)}_{R}(\tilde{\bm{\phi}})\}_{R=1}^\infty$ is a Cauchy sequence, i.e.,
\begin{equation}\label{limit-1}
\lim_{R,F\to\infty}\left|\left|\E_{h,t}\hat{G}^{(h,t)}_{R}(\tilde{\bm{\phi}})-\E_{h,t}\hat{G}^{(h,t)}_{F}(\tilde{\bm{\phi}})\right|\right| = 0,
\end{equation}
and ii)
\begin{equation}\label{limit-2}
\lim_{F\to\infty}\left|\left|\E_{h,t}\hat{G}^{(h,t)}_{F}(\tilde{\bm{\phi}})-\E_{h,0}\hat{G}^{(h,0)}_{F}(\tilde{\bm{\phi}})\right|\right| = 0.
\end{equation}
Equation \ref{limit-1} and equation \ref{limit-2} imply that we may use $\hat{G}^{(h,0)}_{F}(\bm{\phi}^{(h,0)})$, i.e. the one-sample Monte Carlo estimate of $\E_{h,0}\hat{G}^{(h,0)}_{F}(\bm{\phi}^{(h,0)})$, as an estimate of $\E_{h,t}\hat{G}^{(h,t)}_{R}(\bm{\phi}^{(h,0)})$, when  $F\gg R$.
To distinguish between the two sample sizes, we call $R$ the iteration sample size and $F$ the epoch sample size.
According to Section 9.3 in \citet{mcbook}, the standard deviation of $\hat{G}^{(h,0)}_{F}(\bm{\phi}^{(h,0)})$ is of order $O(1/\sqrt{F})$, which means this one-sample Monte Carlo estimate is credible when the epoch sample size $F$ is large. 
An advantage of this estimate is that, once we have computed $\hat{G}^{(h,0)}_{F}(\bm{\phi}^{(h,0)})$ at the beginning of epoch $h$, we can hold it as a constant during the whole epoch.
Finally, the resulting variance reduced reweighted wake-sleep estimator (RWSVR) is
\begin{equation}\label{rwsvr-estimator}
 \hat{G}_{\mathrm{rwsvr}}(\bm{\phi}^{(h,t)}) = \hat{G}^{(h,t)}_R(\bm{\phi}^{(h,t)}) - \hat{G}^{(h,t)}_R(\bm{\phi}^{(h,0)}) + \hat{G}^{(h,0)}_{F}(\bm{\phi}^{(h,0)}),
\end{equation}
where $\hat{G}^{(h,t)}_R(\bm{\phi}^{(h,t)})$ and $\hat{G}^{(h,t)}_R(\bm{\phi}^{(h,0)})$ are calculated based on the same sample $\{(\tau^i, \bm{l}^i)\}_{i=1}^R\overset{\mathrm{i.i.d.}}{\sim} q_{\bm{\phi}^{(h,t)}, \bm{\psi}^{(h,t)}}(\tau,\bm{l})$.

Similar to the original RWS estimator, the proposed RWSVR estimator is biased when the iteration sample size $R$ is finite.
However, the RWSVR estimator $\hat{G}_{\mathrm{rwsvr}}(\bm{\phi}^{(h,t)})$ is strongly consistent (see Definition 2.10 in \citet{shao2003mathematical}) as an estimator of $\hat{G}^{(h,t)}_R(\bm{\phi}^{(h,t)})$ when $R,F\to\infty$ and its variance is substantially smaller than that of the RWS estimator as the algorithm converges if $F\gg R$, as we summarize in the following theorem.

\begin{theorem}\label{thm-rwsvr}
Suppose that i) $\forall \tau, \bm{l}$ and $\bm{\psi}$, $q_{\bm{\phi}, \bm{\psi}}(\tau,\bm{l}) \in C^1(\mathbb{R}^{|p|})$ as a function of CPDs $p$; ii) $G(\phi)$ is $L_G$-Lipschitz continuous and with probability one, $\hat{G}^{(h,t)}_R(\phi)$ is $L_G$-Lipschitz continuous; ii) $\lim_{h\to\infty} \sup_t\mathbb{E}||\bm{\phi}^{(h,t)}-\bm{\phi}^\ast||^2=0$ and $\lim_{h\to\infty} \sup_t\mathbb{E}||\bm{\psi}^{(h,t)}-\bm{\psi}^\ast||^2=0$ for some point $\bm{\phi}^\ast$ and $\bm{\psi}^\ast$.
Then 
\begin{enumerate}
    \item $\hat{G}_{\mathrm{rwsvr}}(\bm{\phi}^{(h,t)})\overset{\mathrm{a.s.}}{\longrightarrow} G(\bm{\phi}^{(h,t)})$ as $R,F\to\infty$;
    \item $\lim_{h\to\infty} \sup_t\mathbb{E}||\hat{G}_{\mathrm{rwsvr}}(\bm{\phi}^{(h,t)})-G(\bm{\phi}^{(h,t)})||^2\overset{\circ}{\approx}O(1/F)$.
\end{enumerate}
\end{theorem}

A proof of Theorem \ref{thm-rwsvr} is provided in Appendix \ref{pf-thm-rwsvr}.
For gradient estimates of the branch length parameters, we follow \citet{zhang2018vbpi} and use the reparameterization trick which leads to the following gradient estimator
\begin{equation}\label{reparam-estimator}
\hat{H}_{\mathrm{reparam}}(\bm{\phi}^{(h,t)},\bm{\psi}^{(h,t)}) = \sum_{i=1}^{R}\tilde{w}^i(\bm{\phi}^{(h,t)},\bm{\psi}^{(h,t)})\nabla_{\bm{\psi}}\log\left(\frac{p(\tau^i, g_{\bm{\psi}^{(h,t)}}(\bm{\epsilon}^i|\tau^i), \bm{Y})}{q_{\bm{\phi}^{(h,t)}, \bm{\psi}^{(h,t)}}(\tau^i, g_{\bm{\psi}^{(h,t)}}(\bm{\epsilon}^i|\tau^i))}\right)
\end{equation}
where $\{\bm{\epsilon}^i\}_{i=1}^R \overset{\mathrm{i.i.d.}}{\sim} \mathcal{N}(\bm{0}_{|\bm{\psi}|}, \bm{I}_{|\bm{\psi}|})$ and $g_{\bm{\psi}}(\bm{\epsilon}|\tau) = \exp(\bm{\mu}_{\bm{\psi}, \tau} + \bm{\sigma}_{\bm{\psi}, \tau}\odot \bm{\epsilon}).$
See Algorithm \ref{alg-rwsvr} for pseudo-code of VBPI with our proposed RWSVR estimator.
\begin{algorithm}[!t]
\caption{VBPI with The RWSVR Estimator}\label{alg-rwsvr}
\KwIn{Observed sequence data $\bm{Y}$; iteration sample size $R$; epoch sample size $F$; number of iterations per epoch $T$.
}
Initialize $\bm{\phi}^{(0,0)}, \bm{\psi}^{(0,0)}$; $h=0$;
\While{not converge}{
Sample $F$ phylogenetic trees $\{(\tau^i, \bm{l}^i)\}_{i=1}^{F}$ from $q_{\bm{\phi}^{(h,0)}, \bm{\psi}^{(h,0)}}(\tau,\bm{l})$ and calculate $\hat{G}^{(h,0)}_{F}(\bm{\phi}^{(h,0)})$\;
\For{$t=0,\ldots, T-1$}{
Sample $R$ phylogenetic trees $\{(\tau^i, \bm{l}^i)\}_{i=1}^{R}$ from $q_{\bm{\phi}^{(h,t)}, \bm{\psi}^{(h,t)}}(\tau,\bm{l})$ and calculate $\hat{G}^{(h,t)}_{R}(\bm{\phi}^{(h,t)})$,  $\hat{G}^{(h,t)}_{R}(\bm{\phi}^{(h,0)})$\;
Calculate $\hat{G}_{\mathrm{rwsvr}}(\bm{\phi}^{(h,t)})$ and $\hat{H}_{\mathrm{reparam}}(\bm{\phi}^{(h,t)}, \bm{\psi}^{(h,t)})$ according to equation (\ref{rwsvr-estimator}) and (\ref{reparam-estimator}) respectively\;
Update $\bm{\phi}^{(h,t)}, \bm{\psi}^{(h,t)}$ to $\bm{\phi}^{(h,t+1)}, \bm{\psi}^{(h,t+1)}$ respectively using stochastic gradient ascent.
}
$\bm{\phi}^{(h+1,0)} \leftarrow \bm{\phi}^{(h,t)}; \bm{\psi}^{(h+1,0)}\leftarrow \bm{\psi}^{(h,t)}; h\leftarrow h+1$\;
}
\end{algorithm}

\section{Experiments}\label{sec4-experiments}
In this section, we evaluate the effectiveness and efficiency of the proposed improved techniques for SBN-based tree topology inference. We focus on two common tasks: tree topology probability estimation and variational Bayesian phylogenetic inference.
For tree topology probability estimation, we first compare different methods (with and without variance reduction) on several synthetic data sets.
We then apply all proposed and previous baseline methods on eight real data sets for tree posterior estimation based on MCMC samples. These data sets, which we will call DS1-8, consist of sequences from 27 to 64 eukaryote species with 378 to 2520 site observations (see Table \ref{table-exp-sbn}).
We use the inclusive KL divergence from the estimated distributions $q_{\bm{\phi}}(\tau)$ to the target distributions $p(\tau|\boldsymbol{Y})$ (estimated from extremely long MCMC runs) to measure the approximation accuracy of different methods.
When empirically evaluating the inclusive KL divergence $\mathrm{KL}(p(\tau|\boldsymbol{Y})|q_{\bm{\phi}}(\tau))$, a tree topology $\tau$ from the empirical sample of $p(\tau|\boldsymbol{Y})$ may not be contained in the support of $q_{\bm{\phi}}(\tau)$ and KL would not be properly defined. To address this numerical issue, all the tree probability estimates of $q_{\bm{\phi}}(\tau)$ are clipped to $10^{-40}$ before computing the KL divergence.
A similar strategy is also considered in \citet{zhang2022variational}.
For simplicity, we will use ``KL divergence'' for inclusive KL divergence unless otherwise specified.
For SEMVR, the clipping value is set to be $\lambda\approx 2.22\times 10^{-16}$, which is the machine precision of 64-bit floating point numbers.
For variational Bayesian phylogenetic inference, we compare the proposed RWSVR estimator with the RWS estimator and the VIMCO estimator.
We examine the performances of different gradient estimators on a challenging synthetic data set and six of the aforementioned real data sets: DS1-4,7-8.\footnote{We omit DS5 and DS6 here as the posteriors on these two data sets are extremely diffuse.}
The KL divergence to the target distributions, lower bound, and marginal likelihood estimates from different methods are reported for comparison.

\subsection{Tree Topology Probability Estimation}\label{sec-exp-sbn}
We first conduct experiments on a simulated setup to empirically investigate the performance of the proposed stochastic algorithms on estimating tree topology probabilities. 
Following \citet{zhang2018sbn}, we choose a tractable but challenging tree topology space, i.e., the space of unrooted tree topologies with 8 leaves, which contains 10395 unique tree topologies. These tree topologies are given an arbitrary order.
To investigate the approximation performance on targets of different degrees of diffusion, we generate target distributions by drawing samples from the symmetric Dirichlet distributions $\operatorname{Dir}(\beta \bm{1})$ of order 10395 with a variety of concentration parameter $\beta$s. The target distribution becomes more diffuse as $\beta$ increases. Simulated data sets are then obtained by selecting tree topologies with the top $K$ largest probabilities.
The resulting probability estimation is challenging in that the target probabilities of the tree topologies are assigned regardless of the similarity among them. 
We initialize CPDs (or the latent parameters of CPDs for SGA and SVRG) using the maximum simple average lower bound estimates as in \citet{zhang2018sbn}.
We set the minibatch size $B=1$ in all experiments. 
We set the learning rate $\rho$ to be $0.001$ for SEM, $0.01$ for SEMVR, $0.0001$ for SGA, and $0.001$ for SVRG.
For SEM and SGA, we use a decreasing learning rate schedule with a decay rate of 0.75 every 50 epochs, i.e., $\rho_n=\rho (0.75)^{[n/(50T)]}$, where $T$ is the number of iterations per epoch.
The number of iterations per epoch is set to be $T=1000$ in all settings.
We run each method 200,000 iterations to ensure convergence.
We vary $\beta$ and $K$ to control the difficulty of the learning task, and the results for each configuration are averaged over 10 independent runs. 
Since the target distributions are known, we use KL divergence from the estimated distributions to the target distributions to measure the approximation accuracy of different methods. 

\begin{figure}[t!]
\includegraphics[width=\textwidth]{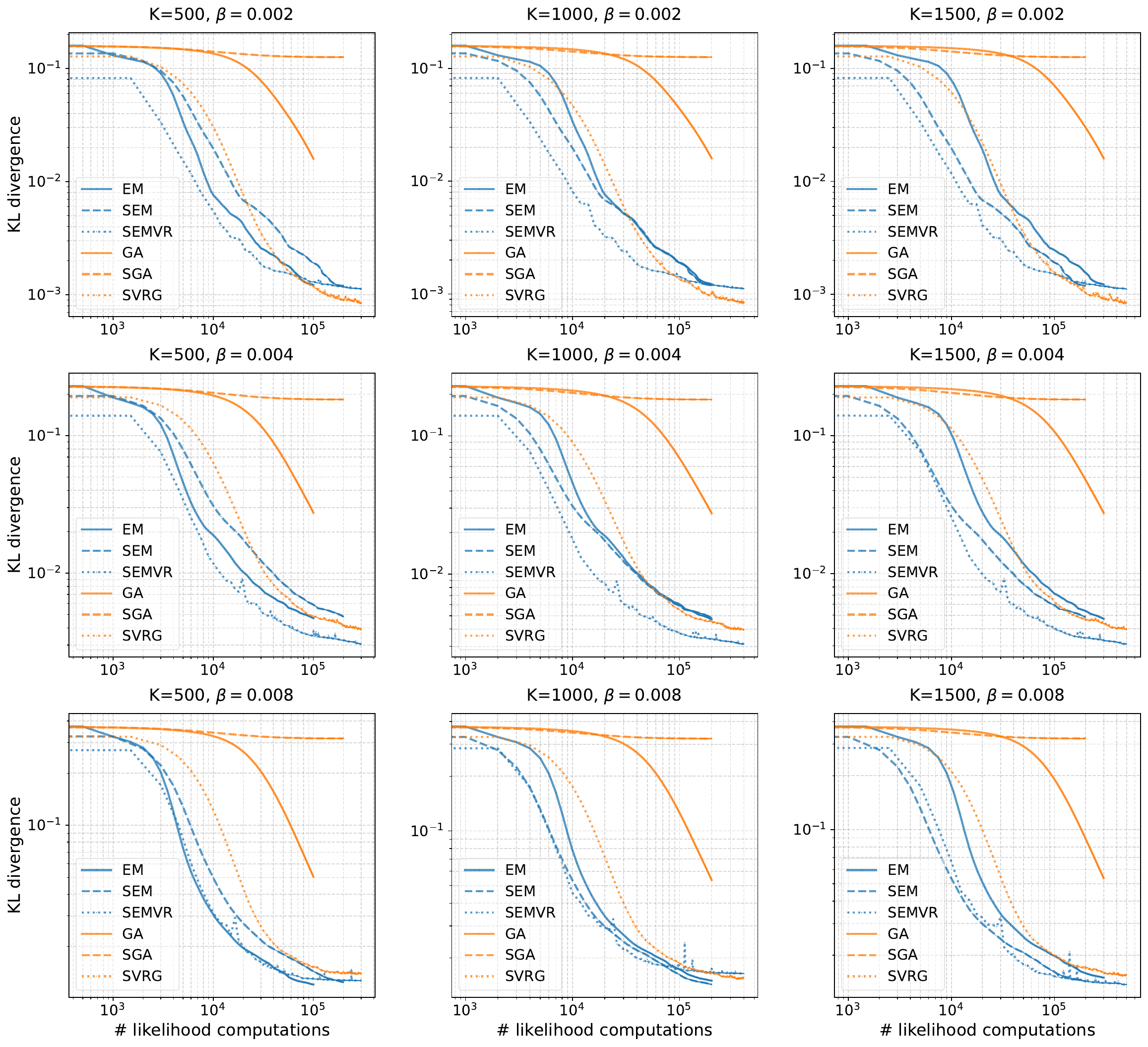}
\caption{Performance on a challenging tree probability estimation problem with simulated data using varying $\beta$ and $K$.
The `\# likelihood computations' refers to the number of likelihood computations during training. The results are averaged over 10 independent runs.
}
\label{fig-sbn-simu-ext}
\end{figure}

Figure \ref{fig-sbn-simu-ext} shows the KL divergence of different methods as a function of the number of likelihood computations over a variety of configurations of the concentration parameter $\beta$ and sample size $K$. 
We see that both SEM and SEMVR converge faster than the full batch EM algorithm when the sample size $K$ is large.
We also considered optimizing SBNs using the full-batch gradient with a learning rate of 0.01, called gradient ascent (GA), as an additional baseline.
When equipped with variance reduction, SEMVR consistently enjoys faster convergence speed than SEM, especially when $\beta$ is small (the target distribution becomes less diffuse).
As for the final results, SEMVR reaches lower KL divergence than EM and SEM when $\beta$ becomes smaller.
Although SGA performs poorly, SVRG improves upon SGA by a large margin in both speed and approximation accuracy and performs comparably to EM-based methods.
With a full-batch gradient, the convergence speed of GA is also much slower than that of other methods except SGA.
These results show that variance reduction can be helpful for accelerating convergence while providing on par or better approximation performance. 

We also test our methods on large unrooted tree topology space posterior estimation on $8$ real data sets, DS1-8, that are commonly used to benchmark phylogenetic MCMC methods \citep{Lakner08, hohna2012guided, larget2013estimation, whidden2015quantifying}. For each of these data sets, 10 single-chain MrBayes \citep{Ronquist12} replicates are run for one billion iterations and sampled every 1000 iterations, with the first 25\% discarded as burn-in for a total of 7.5 million posterior samples per data set.
For all MCMC runs, we assume a uniform prior on the tree topology, an i.i.d. exponential prior (Exp(10)) for the branch lengths and the simple Jukes and Cantor \citep{jukes1969evolution} substitution model.
These extremely long ``golden runs'' form the ground truth to which
we will compare various posterior estimates based on standard runs.

For the standard runs, we follow \citet{zhang2018sbn} and run MrBayes on each data set with 10 replicates of 4 chains and 8 runs until the runs have ASDSF (the standard convergence criteria used in MrBayes) less than 0.01 or a maximum of 100 million iterations. The posterior samples are collected every 100 iterations of these runs and the first 25\% are discarded as burn-in.
We test the stochastic algorithms SEM and SGA, together with their variance reduced counterparts SEMVR and SVRG on posterior estimation based on those MCMC samples in each of the 10 replicates for each data set.
We initialize CPDs (or the latent parameters of CPDs for SGA and SVRG) using the maximum simple average lower bound estimates as in \citet{zhang2018sbn}.
The minibatch size $B$ is set to be 1 for all stochastic algorithms, and we find it works well.
The number of iterations per epoch is fixed to be $T=1000$ for all data set.
We also include an additional baseline: the simple average (SA) of empirical frequencies introduced in \citet{zhang2018sbn}, which is defined as the empirical frequency of CPDs without any further refinement.
The learning rate is set to be $0.01$ for SEMVR, $0.001$ for SEM and SVRG, and $0.0001$ for SGA.
Please check Appendix \ref{app-lr} for an ablation study on the learning rates.
For SEM and SGA, we use a decreasing learning rate schedule with a decay rate of 0.75 every 50 epochs, i.e., $\rho_n=\rho (0.75)^{[n/(50T)]}$, where $T$ is the number of iterations per epoch.
We set $\alpha=0.0001$ for EM-$\alpha$, SEM-$\alpha$, and SEMVR-$\alpha$, following \citet{zhang2018sbn}.
Results are collected after 300 epochs or when the change between the log-likelihoods of two successive epochs is less than $10^{-5}$.

\begin{figure}[t]
    \centering
\includegraphics[width=\linewidth]{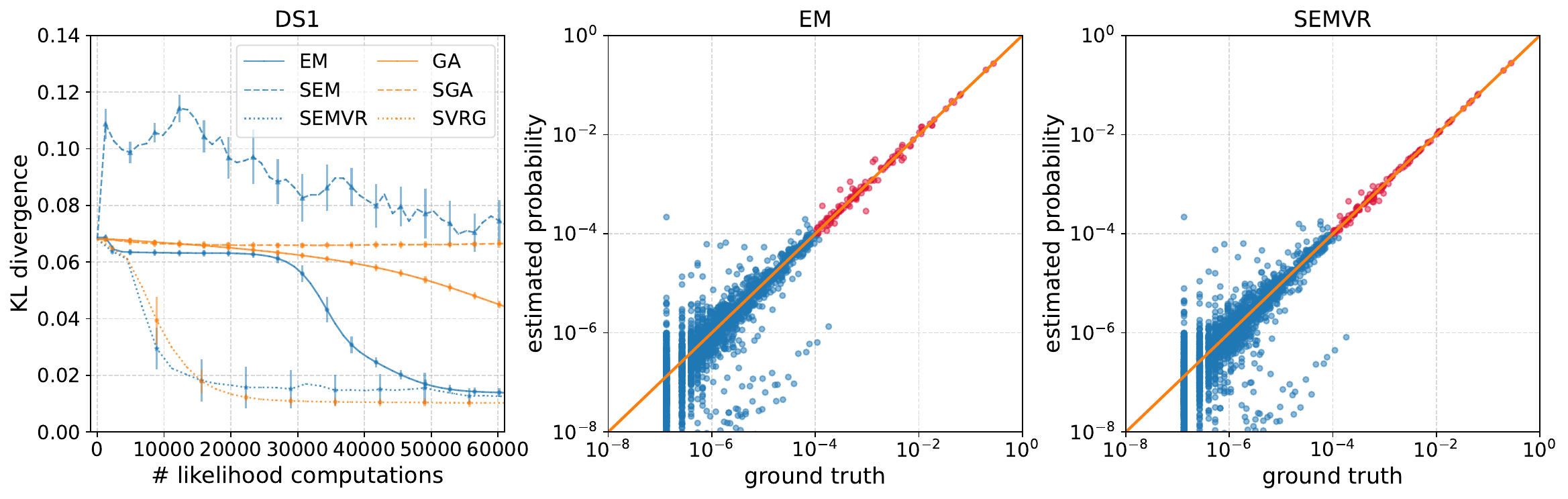}
\caption{Comparison on DS1. \textbf{Left:} KL divergence between estimated posterior probabilities and the ground truth during training.
The `\# likelihood computations' refers to the number of likelihood computations during training. The results are averaged over 10 replicates and the error bars show one standard deviation. \textbf{Middle:} The posterior probabilities estimated with EM v.s. the ground truth. \textbf{Right:} The posterior probabilities estimated with SEMVR v.s. the ground truth.}
\label{fig-sbn-training}
\end{figure}

Figure \ref{fig-sbn-training} shows the performance of different algorithms on DS1.
The left plot shows the KL divergence between the estimated posteriors and the ground truth as a function of the number of likelihood computations.
Note that the KL divergence is evaluated every epoch and the number of likelihood computations per epoch can depend on the methods (Table \ref{tab:computation-complexity}), resulting in unequal intervals between error bars.
The GA method with a learning rate of 0.01 is also considered here.
We see that both EM and GA converge slowly at the beginning (likely being trapped at a stationary point).
This validates our previous claim that EM  may suffer from slow convergence and get trapped at local modes or stationary points.
Both SEM and SGA suffer from large variances and the KL divergence cannot get down even if small step sizes are used (SEM eventually arrives at a smaller KL divergence when trained longer, see Table \ref{table-exp-sbn}).
In contrast, for their variance reduced variants, SEMVR and SVRG, the KL divergences decrease fast and almost converge after 20,000 likelihood computations,
albeit being slightly unstable at the beginning due to the relatively larger learning rates.
Not only helpful for improving the computation efficiency, stochastic optimization may also be helpful for improving the probability estimation of tree topologies due to its exploration capability.
The middle and right plots compare EM and SEMVR estimates with the ground truth.
We see that SEMVR can provide more accurate posterior estimates for those tree topologies with high posterior probabilities.
When applied to a broad range of data sets, we find that SEMVR and SVRG tend to provide better posterior estimates than the other algorithms (Table \ref{table-exp-sbn}).
With regularization, SEMVR-$\alpha$ performs better than SEMVR in all cases.

\begin{table}[t]
\centering
\caption{ KL divergence between SBN-based posterior estimates and the ground truth. The number of sampled trees means the number of unique trees in the standard runs, which reflects the dispersion of the posterior distribution. The results are averaged over 10 replicates.}
\label{table-exp-sbn}
\centering
\setlength{\tabcolsep}{0.1cm}{}
\begin{tabular*}{\textwidth}{@{\extracolsep\fill}ccccccccc}
\toprule
data set & DS1 & DS2 & DS3 & DS4 & DS5 & DS6 &DS7 & DS8\\
\midrule
\# taxa & 27 & 29 & 36 & 41 & 50 & 50 & 59 & 64 \\
\# sampled trees & 1228 & 7 & 43 & 828 & 33752 & 35407 & 1125 & 3067 \\
\midrule
SRF & 0.0155  & 0.0122 & 0.3539 & 0.5322 & 11.5746 & 10.0159 & 1.2765 & 2.1653\\
SA & 0.0687 &0.0218&0.1152&0.1021&0.8952 &0.2613& 0.2341&0.2212\\
\midrule 
EM & 0.0136  &0.0199 &0.1243 &0.0763 &0.8599 &0.3016  &0.0483  &0.1415\\
SEM &0.0366 &0.0131 &0.1117 &0.0903 &0.9210 &0.3350 &0.0549 &0.1714\\
SGA & 0.0666&0.0215& 0.1161&0.1044&0.9083 &\bf{0.2667}& 0.2345&0.2275\\
SEMVR & 0.0125 &0.0157 &0.1229  &0.0793 &0.8364 &0.3017&0.0403&0.1428 \\
SVRG & \bf{0.0088}& \bf{0.0120}& \bf{0.1003}&\bf{0.0671} &\bf{0.8172}&0.2817 &\bf{0.0360}&\bf{0.1234}\\
\midrule
EM-$\alpha$ & 0.0130 &0.0128 &\bf{0.0882} &\bf{0.0637} &0.8218&0.2786&0.0399&0.1236 \\
SEM-$\alpha$ &0.0307 &0.0127 &0.0891 &0.0752 &0.8717 &0.3096 &0.0519 &0.1494 \\
SEMVR-$\alpha$ &  \bf{0.0100}& \bf{0.0120}& 0.0918& 0.0649& \bf{0.8176}& \bf{0.2778}&\bf{0.0377} & \bf{0.1197}\\ 
\bottomrule
\end{tabular*}
\end{table}

\subsection{Variational Bayesian Phylogenetic Inference}
We now investigate the performance of the RWSVR estimator for learning SBN-based variational distributions on phylogenetic trees under the variational Bayesian phylogenetic inference framework.
We use Adam \citep{ADAM} and AMSGrad \citep{AMSGRAD} for stochastic gradient ascent.
Results are collected after 200,000 parameter updates.

As before, we first conduct experiments on a simulated setup, using the same space of unrooted phylogenetic tree topologies with 8 leaves without branch lengths and the same target distribution $p_{0}(\tau)$ generated from the symmetric Dirichlet distributions $\operatorname{Dir}(\beta \bm{1})$.
Following \cite{zhang2018vbpi}, we used $\beta=0.008$ to provide enough information for inference while allowing for adequate diffusion in the target.
Note that there are no branch lengths in this simulated model and the evidence lower bound (ELBO) is
\begin{equation}
L(\phi)=\mathbb{E}_{Q_{\phi}(\tau)} \log \left(\frac{p_{0}(\tau)}{Q_{\bm{\phi}}(\tau)}\right) \leq 0
\end{equation}
with the exact evidence being $\log (1)=0$.
The CPDs in the variational distribution $Q_{\bm{\phi}}(\tau)$ are uniformly initialized, i.e., all the entries in $\bm{\phi}$ are initialized as zeros.
We use both the RWS and RWSVR estimators with iteration sample size $R=10$ and $R=20$, and fix the epoch sample size as $F=1000$.
Note that in RWS and RWSVR, the objective function is the expected log-likelihood of $Q_{\bm{\phi}}$ instead of the ELBO, and we used self-normalized importance sampling for gradient estimation, where samples from $Q_{\bm{\phi}}$ that are not in the support of $p_{0}$ have zero weights and would not cause numerical instability.
The number of iterations per epoch is set to be $T=100$.
We use a learning rate of 0.002 in AMSGrad for RWS and RWSVR, with a decay rate of 0.75 every 20,000 iterations.

\begin{figure}
\centering
\includegraphics[width=0.75\textwidth]{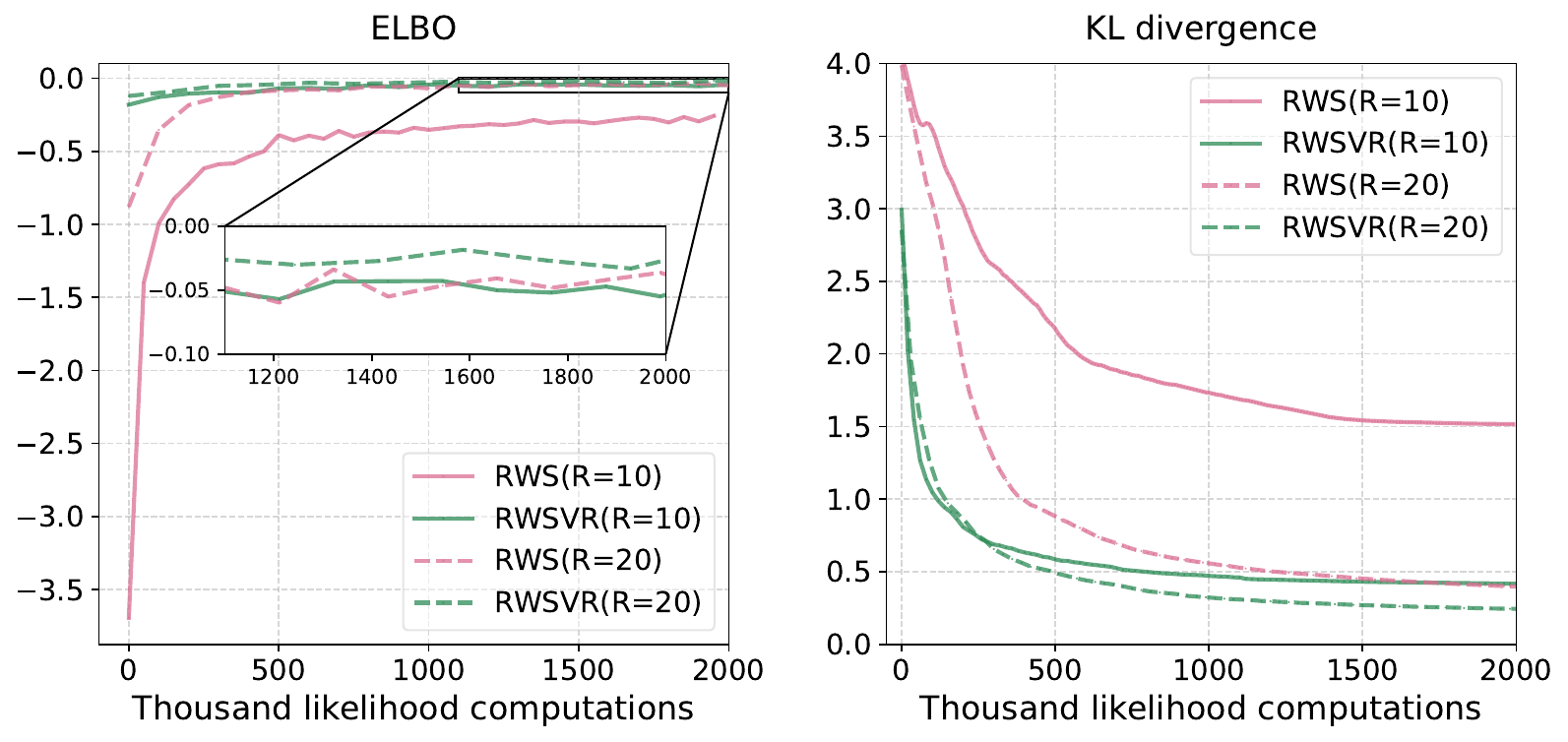}
\caption{
ELBO and KL divergence as a function of the number of likelihood computations on a synthetic data set of unrooted phylogenetic trees with 8 leaves, varying the number of particles in RWS and RWSVR.
}
\label{fig-vbpi-simu}
\end{figure}

Figure \ref{fig-vbpi-simu} depicts the resulting ELBO and KL divergence to the ground truth as a function of the number of likelihood computations.
We see that in both cases ($R=10,20$), RWSVR converges faster and tends to provide a higher lower bound than RWS, especially when the iteration sample size $R$ is small. 
The evolution of KL divergence is consistent with the ELBO.
The fast start of RWSVR is partly due to variance reduction that provides a more stable optimization direction in the beginning phase, similar to the tree topology probability estimation tasks.
Moreover, a large epoch sample size also introduces extra variability which allows RWSVR to jump out of local minima and acquire more accurate tree topology probability estimates when the iteration sample size $R$ is small. 

\begin{table}[t]
\tiny
\caption{KL divergence to the ground truth, evidence lower bound (ELBO) and marginal likelihood (ML) estimates of different methods across 6 benchmark data sets for Bayesian phylogenetic inference.
The ``\# GT trees'' refers to the number of tree topologies in the ground truth which reflect the concentration of the posterior distribution.
The ``\# supp trees'' refers to the number of rooted tree topologies in the support of $q_{\bm{\phi}}(\tau)$.
The results of KL divergence are averaged over 10 independent trainings with standard deviation in brackets. 
The marginal likelihood estimates of all variational methods are obtained via importance sampling using 1000 samples. The results of ELBO and ML are averaged over 100 independent runs with standard deviation in brackets.
}
\label{table-exp-vbpi}
\centering
\setlength{\tabcolsep}{0.03cm}{}
\begin{tabular*}{\textwidth}{@{\extracolsep\fill}cccccccc}
\toprule
\multicolumn{2}{c}{data set}                                & DS1               & DS2                & DS3                & DS4       & DS7 & DS8         \\
\multicolumn{2}{c}{\# taxa}                                 & 27                & 29                 & 36                 & 41          & 59 & 64       \\
\multicolumn{2}{c}{\# GT trees}&{2784}&{42}&{351}&{11505}&{11525}&{82162}\\
\multicolumn{2}{c}{\# supp trees}&{$1.14\times 10^{10}$}&{$2.46\times 10^8$}&{$1.01\times 10^{11}$}&{$4.88\times 10^{11}$}&{$1.66\times 10^{15}$}&{$1.67\times 10^{19}$}\\
\multicolumn{2}{c}{\# parameters}&{8235}& {4349} & {4572} & {8624} &{4322}&{12042}\\
\midrule
& VIMCO & 0.0741(0.001)& 0.0197(0.001)& 0.0802(0.001) & 0.1011(0.003)&0.2170(0.028) &  \textbf{0.4694(0.064)}\\
& RWS& 0.0803(0.002)&0.0113(0.000)& 0.0706(0.009)& 0.1652(0.006) & 0.3080(0.038)&0.7570(0.058)\\
\rot{\rlap{~KL}} & RWSVR & \textbf{0.0438(0.014)}  & \textbf{0.0006(0.001)}   & \textbf{0.0085(0.001)}   & \textbf{0.0461(0.005)} & \textbf{0.0816(0.022)} & 0.5054(0.221)   \\
\midrule
    & VIMCO & -7111.40(9.380)          & -26369.51(0.755)          & -33736.64(0.326)          & -13332.47(0.645)     & -37335.18(0.128) & -8655.53(0.427)     \\
  & RWS   & -7110.36(0.326)          & -26368.82(0.054)          & -33736.27(0.056)          & \textbf{-13331.96(0.136)} & -37335.12(0.128)&-8655.39(0.230)\\
      & RWSVR & \textbf{-7110.28(0.114)} & \textbf{-26368.78(0.054)} & \textbf{-33736.21(0.053)} & -13332.02(0.108)  & \textbf{-37335.11(0.126)}   & \textbf{-8655.33(0.274)}     \\ \midrule
\rot{\rlap{~~~~ELBO}}& VIMCO & -7108.41(0.192) & -26367.71(0.089)          & -33735.10(0.103)          & -13329.96(0.234) & -37332.00(0.327)     & -8650.68(0.541)     \\
\multicolumn{1}{c}{}                              & RWS   & -7108.42(0.188) & -26367.71(0.089)          & -33735.10(0.109)          & -13329.97(0.234) & -37332.01(0.357)    & -8650.71(0.518)      \\
\rot{\rlap{~ML}}                             & RWSVR & \textbf{-7108.42(0.174)}          & \textbf{-26367.71(0.084)} & \textbf{-33735.09(0.090)} & \textbf{-13329.95(0.216)} & \textbf{-37331.98(0.325)} & \textbf{-8650.67(0.508)} \\
\bottomrule
\end{tabular*}
\end{table}

Next, we evaluate the proposed RWSVR gradient estimator for phylogenetic posterior estimation via VBPI on real data sets.
The ground truth posterior estimates are formed based on extremely long golden runs as described in section \ref{sec-exp-sbn}, and the numbers of tree topologies in the ground truth are reported in Table \ref{table-exp-vbpi}.
We conduct experiments on DS1-4,7-8 where the posteriors are relatively less diffuse and the ground truth tree topology posterior probabilities can be more reliably estimated from MCMC runs.\footnote{
{
Here, we omitted DS5 and DS6 as the posterior distributions for these two datasets are much more diffuse than the other datasets (the number of unique tree topologies in the ground truth is 1516877 for DS5 and 809765 for DS6). As a result, a much larger epoch sample size $F$ is required for a more accurate RWSVR gradient estimator that may hinder computational efficiency.
}}
We gather the support of CPDs from 10 replicates of 10000 ultrafast maximum likelihood bootstrap trees \citep{minh2013ultrafast}.
The CPDs in the variational distribution $Q_{\bm{\phi}}(\tau)$ are uniformly initialized, i.e., all the entries in $\bm{\phi}$ are initialized as zeros.
Following \citet{rezende2015variational}, we use an annealed likelihood $\left[p\left(\boldsymbol{Y} |\tau^i, \boldsymbol{q}^i\right)\right]^{\beta_t}$ in the training objectives, where $\beta_t \in[0,1]$ is an inverse temperature that follows a schedule $\beta_t=\min (1, 0.001 + t / 100000)$, going from $0.001$ to 1 after 99900 iterations. 
For DS1-4 and DS7, we set the epoch sample size $F=1000$; for DS8, we use a larger epoch sample size $F=3000$ as DS8 is more diffuse than the other data sets (Table \ref{table-exp-vbpi}).
The number of iterations per epoch is set to $T=100$ for all data sets.
We use Adam with a learning rate of 0.001 and a decay rate of 0.75 every 20,000 iterations to train the variational approximations using VIMCO, RWS, and RWSVR estimators with iteration sample size $R=10$.

\begin{figure}
\centering
\includegraphics[width=0.75\textwidth]{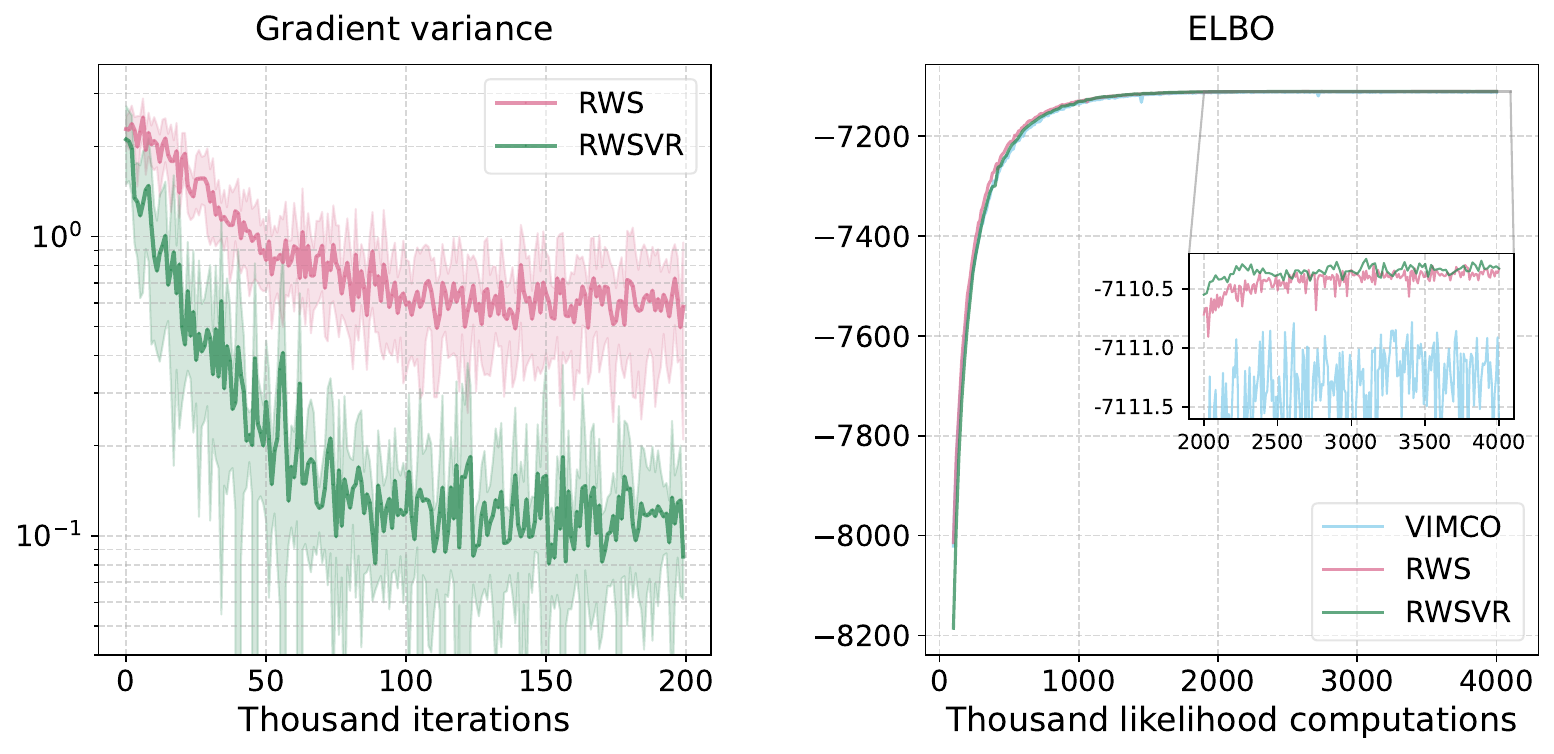}
\caption{
The Log-scaled gradient variance and evidence lower bound (ELBO) as a function of iterations or likelihood computations given by different methods on the real data set DS1.
}
\label{fig-vbpi-ds1}
\end{figure}

\begin{figure}[t]
\centering
\begin{minipage}{0.48\textwidth}
\includegraphics[width=\textwidth]{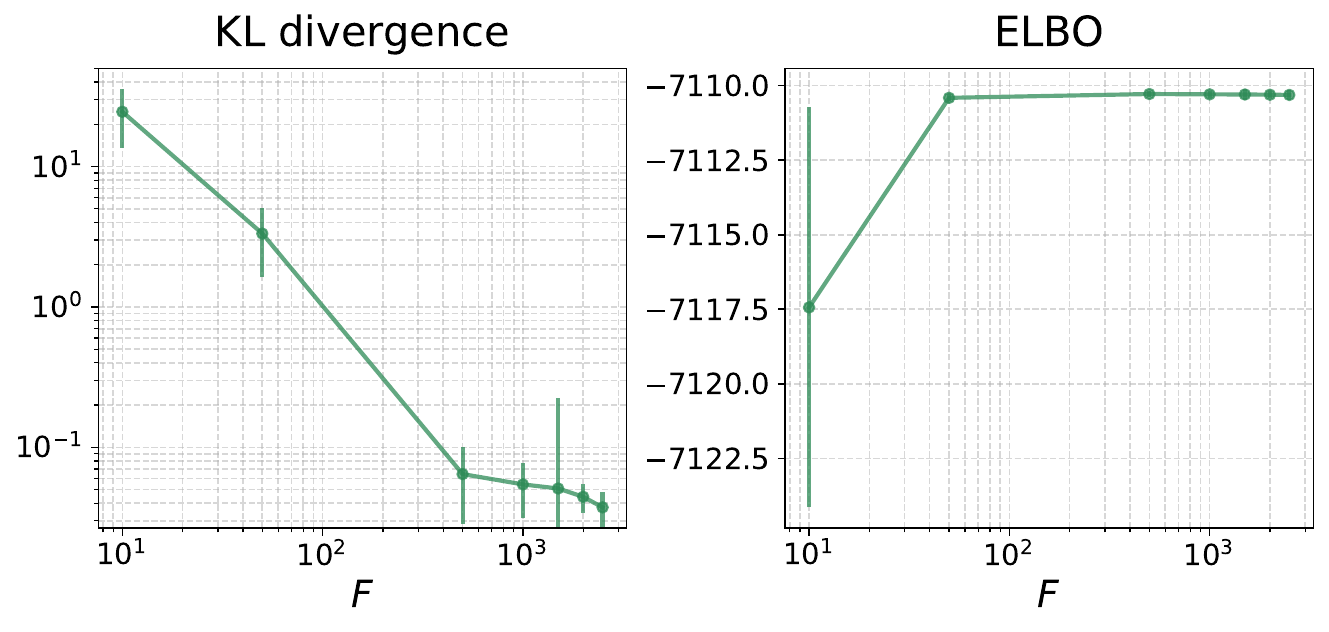}
\end{minipage}
\begin{minipage}{0.48\textwidth}
\includegraphics[width=\textwidth]{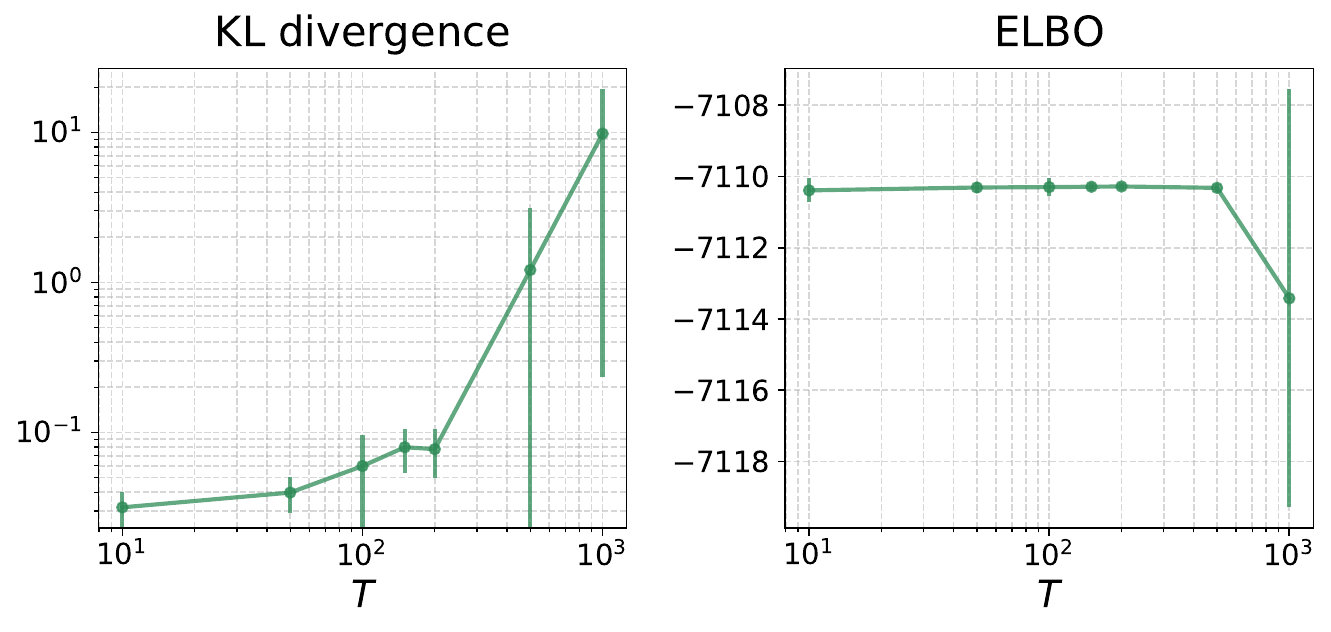}
\end{minipage}
\caption{Ablation studies on DS1. \textbf{Left:} The performance of the RWSVR estimator when varying the epoch sample size ($F$). \textbf{Right:} The performance of the RWSVR estimator when varying the number of iterations trained in each epoch ($T$).
The error bars show one standard deviation over 10 independent runs.}
\label{fig-ablation}
\end{figure}

Table \ref{table-exp-vbpi} shows the resulting KL divergence, ELBO, and marginal likelihood estimates for all gradient estimators on DS1-4,7-8. 
We see that the KL divergences for RWSVR are lower than those for RWS on all 6 real data sets, and they are also lower than those for VIMCO except DS8.
The superiority of RWSVR over RWS can be explained by the behavior of the variances of the resulting gradient estimates.
The left plot in Figure \ref{fig-vbpi-ds1} shows the gradient variance against the number of iterations for both RWSVR and RWS on DS1.
We see that compared to RWS, the variance of the gradient estimates provided by RWSVR has been significantly reduced throughout the entire training process.
The improved training of SBNs can also be helpful for the overall variational approximation, as evidenced by the increased ELBOs and the reduced variance of the marginal likelihood estimates.
We also investigate the effect of variance reduction on training.
The right plot in Figure \ref{fig-vbpi-ds1} shows the ELBO for different methods as a function of the number of likelihood computations.
We find that although RWSVR slightly lags behind RWS and VIMCO at the begining, it surpasses them after around $10^6$ likelihood computations and finally reaches a better ELBO.
Finally, we perform an ablation study on DS1 to investigate the effect of hyperparameter choice on the performance of RWSVR, with different epoch sample size $F$ and number of iterations per epoch $T$. 
The two plots on the left side of Figure \ref{fig-ablation} show the resulting KL divergence and ELBO as a function of $F$ when $T$ is fixed at $100$.
We see that as $F$ increases, the KL divergence becomes lower and the evidence lower bound gets larger.
Therefore, one can expect more accurate posterior estimates from larger epoch sample sizes which is due to the variance reduction effect.
Moreover, this benefit of variance reduction quickly reaches a plateau so that a moderate $F$ (around 500 in this case) would be good enough to make RWSVR perform well on DS1.
The two plots on the right side of Figure \ref{fig-ablation} show the results as a function of $T$ when $F$ is fixed at $1000$.
We can see that the performance of RWSVR stays the same at the beginning, indicating that variance reduction remains effective when $T$ is relatively small.
For large $T$ (more than 200 in this case), RWSVR deteriorates dramatically, which is due to the enlarging difference between $\bm{\phi}^{(h,t)}$ and $\bm{\phi}^{(h,0)}$ as $t$ increases.
Note that the number of likelihood computations during training is $I_{\mathrm{total}}\left(R+F/T\right)$, where $I_{\mathrm{total}}$ is the total number of training iterations.
The choice of $T$, therefore, should strike a good balance between computation efficiency and the effectiveness of large sample expectation estimation for variance reduction. 

\section{Discussion}\label{sec:conclusion}
In this work, we introduced several advanced techniques for phylogenetic tree topology inference based on subsplit Bayesian networks (SBNs). 
We showed that stochastic algorithms can be easily adapted to learn SBNs that scale up to large data sets.
Moreover, variance reduction techniques can also be leveraged to further improve the computational efficiency.
For tree topology probability estimation, we showed that the SEMVR and SVRG algorithms can be adapted which significantly accelerates the previous full batch EM baseline and tends to provide more accurate estimates as well.
For variational Bayesian phylogenetic inference, we proposed the RWSVR estimator which can provide gradient estimates with substantially smaller variance, and thus significantly outperformed the previous RWS estimator, especially in terms of tree topology posterior estimation.
Extensive synthetic and real data experiments have demonstrated the effectiveness and efficiency of our proposed algorithms. 
One limitation of the RWSVR estimator is that a small epoch sample size can result in inaccurate gradient estimation on extremely diffuse data sets, invoking the need for a large epoch sample size that may hinder the computational efficiency.
 
There are many opportunities for future investigation.
For tree topology probability estimation, we mainly focus on leaf-labeled bifurcating trees.
As SBN can be easily adapted for general leaf-labeled trees, we may further investigate the performance of our proposed learning techniques for general, multifurcating trees.
For variational Bayesian phylogenetic inference, we may also combine the RWSVR estimator with more expressive branch length distributions \citep{zhang2020vbpinf} for better overall variational approximation.

\section*{Declarations}
\paragraph{Funding}
The research of Cheng Zhang was supported in part by National Natural Science Foundation of China (grant no. 12201014 and grant no. 12292983), National Institute of Health grant AI162611, the National Engineering Laboratory for Big Data Analysis and Applications, the Key Laboratory of Mathematics and Its Applications (LMAM), and the Fundamental Research Funds for the Central Universities, Peking University.
The research of Minghua Deng was supported by the National Key Research and Development Program of China under Grant 2021YFF1200902 and the National Natural Science Foundation of China under Grant 32270689 and 12126305.
This work was supported by High-performance Computing Platform of Peking University.

\paragraph{Conflict of interest}
The authors declare that they have no conflict of interest.

\paragraph{Data availability}
The data that support the findings of this study are available in the databases of previous publications as detailed in the article.

\paragraph{Code availability}
The code is available at \url{https://github.com/tyuxie/RWSVR}.

\bibliographystyle{nameyear}
\bibliography{sn-bibliography}

\appendix

\section{Details of EM-$\alpha$}\label{app-emalpha}
To improve the generalization ability of EM, one can also assume a Dirichlet prior on CPDs as regularization and then derive the updating scheme. Specifically, if the CPDs $c$ has prior $\mathrm{Dir}(\bm{\alpha}+1)$ with $\bm{\alpha}=\{\alpha_{s|t}\}_{s|t\in\mathbb{S}_{\mathrm{ch|pa}}}\cup \{\alpha_s\}_{s\in\mathbb{S}_{\mathrm{r}}}$, the full-sample Q-function has the form
$$Q^\alpha(c;\hat{c}) = \sum_{s\in\mathbb{S}_{\mathrm{r}}}\left(M_{s}(\hat{c}) +\alpha_s\right)\log c_s + \sum_{s|t\in \mathbb{S}_{\mathrm{ch|pa}}} \left(M_{s|t}(\hat{c})+\alpha_{s|t}\right) \log c_{s|t}.$$
In the regularization setting, the E-step in EM is unchanged and the M-step is
\begin{itemize}
\item M-step (regularization): update the estimates of CPDs as $\hat{c}^{(n+1)} = \Phi(M(\hat{c}^{(n)}) + \bm{\alpha})$.
\end{itemize}
In practice, $\bm{\alpha}$ is commonly set by $\alpha_{s|t}=\alpha m_{s|t}$ and $\alpha_{s}=\alpha m_{s|t}$ where $m_{s|t}$ and $m_s$ are the pseudo-counts of $s|t$ and $s$ in $\mathcal{D}$ and the fixed scaler $\alpha$ is the regularization strength \citep{zhang2018sbn}.
We use EM-$\alpha$ to denote EM with regularization in this paper.

\section{Proof of Theorem \ref{thm-rwsvr}}\label{pf-thm-rwsvr}

We first prove the strong consistency (see Definition 2.10 in \citet{shao2003mathematical}) of the RWSVR estimator. In fact, by the strong law of large numbers, we have 
\[
\lim_{R\to\infty}\frac{1}{R}\sum_{i=1}^R w^i(\bm{\phi}, \bm{\psi}) = \E_{q_{\bm{\phi},\bm{\psi}}(\tau,\bm{l})}\frac{p(\tau,\bm{l},Y)}{q_{\bm{\phi},\bm{\psi}}(\tau,\bm{l})}=p(Y),
\]
\[
\lim_{R\to\infty}\frac{1}{R}\sum_{i=1}^R w^i(\bm{\phi}, \bm{\psi}) \nabla_{\bm{\phi}}\log q_{\tilde{\bm{\phi}}}(\tau^i)=\E_{q_{\bm{\phi},\bm{\psi}}(\tau,\bm{l})}\frac{p(\tau,\bm{l},Y)}{q_{\bm{\phi},\bm{\psi}}(\tau,\bm{l})}\nabla_{\bm{\phi}}\log q_{\tilde{\bm{\phi}}}(\tau)=p(Y)\E_{p({\tau,\bm{l}|Y})}\nabla_{\bm{\phi}}\log q_{\tilde{\bm{\phi}}}(\tau).
\]
Therefore,
\[
\lim_{R\to\infty}\hat{G}^{(h,t)}_{R}(\tilde{\bm{\phi}}) =\lim_{R\to\infty} \frac{\frac{1}{R}\sum_{i=1}^R w^i(\bm{\phi}^{(h,t)}, \bm{\psi}^{(h,t)}) \nabla_{\bm{\phi}}\log q_{\tilde{\bm{\phi}}}(\tau^i)}{\frac{1}{R}\sum_{i=1}^R w^i(\bm{\phi}^{(h,t)}, \bm{\psi}^{(h,t)})}=\E_{p({\tau,\bm{l}|Y})}\nabla_{\bm{\phi}}\log q_{\tilde{\bm{\phi}}}(\tau)=G(\tilde{\bm{\phi}}).
\]

We then estimate the order of the variance of the RWSVR estimator. We have the following estimate for the mean squared error (MSE) of $\hat{G}^{(h,t)}_{F}(\tilde{\bm{\phi}})$:
\begin{align*}
\mathbb{E}||\hat{G}^{(h,t)}_{F}(\tilde{\bm{\phi}})-G(\tilde{\bm{\phi}})||^2&=\E\left(\frac{\frac{1}{F}\sum_{i=1}^{F} w^i(\bm{\phi}^{(h,t)}, \bm{\psi}^{(h,t)}) \nabla_{\bm{\phi}}\log q_{\tilde{\bm{\phi}}}(\tau^i)}{\frac{1}{F}\sum_{j=1}^{F} w^j(\bm{\phi}^{(h,t)}, \bm{\psi}^{(h,t)})}-G(\tilde{\bm{\phi}})\right)^2\\
&\approx \frac{1}{F}\frac{\E_{q_{\bm{\phi}^{(h,t)},\bm{\psi}^{(h,t)}}(\tau,\bm{l})}\left(w(\bm{\phi}^{(h,t)},\bm{\psi}^{(h,t)})\left(\nabla_{\bm{\phi}}\log q_{\tilde{\bm{\phi}}}(\tau)-G(\tilde{\bm{\phi}})\right)\right)^2}{\E_{q_{\bm{\phi}^{(h,t)},\bm{\psi}^{(h,t)}}(\tau,\bm{l})}\left(w(\bm{\phi}^{(h,t)},\bm{\psi}^{(h,t)})\right)^2}\\
&=:\frac{\sigma^2(\bm{\phi}^{(h,t)},\bm{\psi}^{(h,t)})}{F}
\end{align*}
where the $\approx$ refers to asymptotic equivalence as $F\to\infty$.
As the algorithm converges, i.e. $(\bm{\phi}^{(h,t)}, \bm{\psi}^{(h,t)})\to(\bm{\phi}^\ast, \bm{\psi}^\ast)$ for some $\bm{\phi}^\ast, \bm{\psi}^\ast$, the variance of $\hat{G}^{(h,t)}_R(\bm{\phi}^{(h,t)}) - \hat{G}^{(h,t)}_R(\bm{\phi}^{(h,0)})$ vanishes since it converges to 0 almost surely with $R$ fixed. Hence
\begin{align*}
 &\mathbb{E}||\hat{G}_{\mathrm{rwsvr}}(\bm{\phi}^{(h,t)})-G(\bm{\phi}^{(h,t)})||^2 \\
 =& \mathbb{E}||\hat{G}^{(h,t)}_R(\bm{\phi}^{(h,t)}) - \hat{G}^{(h,t)}_R(\bm{\phi}^{(h,0)}) + \hat{G}^{(h,0)}_{F}(\bm{\phi}^{(h,0)})- G(\bm{\phi}^{(h,0)})+G(\bm{\phi}^{(h,0)})-G(\bm{\phi}^{(h,t)})||^2\\
 \leq & 2\mathbb{E}||\hat{G}^{(h,t)}_R(\bm{\phi}^{(h,t)}) - \hat{G}^{(h,t)}_R(\bm{\phi}^{(h,0)})||^2+2\mathbb{E}||\hat{G}^{(h,0)}_{F}(\bm{\phi}^{(h,0)})-G(\bm{\phi}^{(h,0)})||^2+2||G(\bm{\phi}^{(h,0)})-G(\bm{\phi}^{(h,t)})||^2\\
 \overset{\circ}{\leq } & 4L_G^2 \mathbb{E}||\bm{\phi}^{(h,t)}-\bm{\phi}^{(h,0)}||^2 + 2\frac{\sigma^2(\bm{\phi}^{(h,t)},\bm{\psi}^{(h,t)})}{F},
\end{align*}
because $G$ is $L_G$-Lipschitz continuous and with probability one $\hat{G}^{(h,t)}$ is $L_G$-Lipschitz continuous for all $h$ and $t$. 
Using the fact 
$$\lim_{h\to\infty}\sup_t\mathbb{E}||\bm{\phi}^{(h,t)}-\bm{\phi}^{(h,0)}||^2\leq \lim_{h\to\infty}\sup_t\mathbb{E}||\bm{\phi}^{(h,t)}-\bm{\phi}^\ast||^2+\lim_{h\to\infty}\mathbb{E}||\bm{\phi}^{(h,0)}-\bm{\phi}^\ast||^2= 0$$
because of the assumption $\lim_{h\to\infty}\sup_t\mathbb{E}||\bm{\phi}^{(h,t)}-\bm{\phi}^\ast||^2=0$,
Therefore, we conclude that
$$\lim_{h\to\infty}\sup_t \mathbb{E}||\hat{G}_{\mathrm{rwsvr}}(\bm{\phi}^{(h,t)})-G(\bm{\phi}^{(h,t)})||^2 \overset{\circ}{\leq} 2\lim_{h\to\infty}\sup_t \frac{\sigma^2(\bm{\phi}^{(h,t)},\bm{\psi}^{(h,t)})}{F} = O(1/F)$$as the algorithm converges.

\section{Ablation Study on Learning Rates for SBNs}\label{app-lr}
In this section, we provide an ablation study on the learning rate of EM, SEM, SGA, SEMVR, and SVRG, and explain how we select the learning rate in section \ref{sec-exp-sbn}.

We first determine a good learning rate of the two baseline methods (without variance reduction) SEM and SGA by grid search on DS1. Table \ref{tab:kl-sem} and Table \ref{tab:kl-sga} report the KL divergence of SEM/SGA with varying learning rates.
SEM can perform worse using a too-large or too-small learning rate, and a moderate 0.001 performs best in practice.
We see that the SGA can get worse with an ambitious learning rate of 0.0003 due to the large variance nature of stochastic gradient-based methods, suggesting a smaller learning rate empirically (We select 0.0001).

\vspace{1em}
\hspace{-1em}
\begin{minipage}{0.48\linewidth}
\captionof{table}{KL divergence obtained by SEM and SEM-$\alpha$ with varying learning rate on DS1. We select $\rho=0.001$ in our implementation.}
\label{tab:kl-sem}
\resizebox{\linewidth}{!}{
\begin{tabular}{ccccc}
\toprule
learning rate $\rho$ & 0.0001 & 0.0003 & 0.001 & 0.003 \\
\midrule
SEM   & 0.0687 & 0.0415 & 0.0366 & 0.0755 \\
SEM-$\alpha$& 0.0674& 0.0377& 0.0307 & 0.0456 \\
\bottomrule
\end{tabular}
}
\end{minipage}
\begin{minipage}{0.48\linewidth}
\captionof{table}{KL divergence obtained by SGA with varying learning rate on DS1. We select $\rho=0.0001$ in our implementation.}
\label{tab:kl-sga}
\resizebox{\linewidth}{!}{
\begin{tabular}{ccccc}
\toprule
learning rate $\rho$ & 0.00001 & 0.00003 & 0.0001 & 0.0003 \\
\midrule
SGA  & 0.0670 & 0.0660 & 0.0666 & 0.0728 \\
\bottomrule
    \end{tabular}
}
\end{minipage}
\vspace{1em}

Then, we set the learning rate of SEMVR to $10\times$ that of SEM, and the learning rate of SVRG to $10\times$ that of SGA, as \citet{johnson2013svrg} and \citet{chen2018semvr} both suggested a much larger learning rate for the VR methods compared to the no-VR methods. 
Empirically, we found this is a good choice for these two methods (Table \ref{tab:kl-semvr} and Table \ref{tab:kl-svrg}).

\vspace{1em}
\hspace{-1em}
\begin{minipage}{0.48\linewidth}
\captionof{table}{KL divergence obtained by SEMVR and SEMVR-$\alpha$ with varying learning rate on DS1. We select $\rho=0.01$ in our implementation.}
\label{tab:kl-semvr}
\resizebox{\linewidth}{!}{
\begin{tabular}{ccccc}
\toprule
learning rate $\rho$ & 0.001 & 0.003 & 0.01 & 0.03 \\
\midrule
SEMVR   & 0.0136 & 0.0135 & 0.0125 & 0.0146 \\
SEMVR-$\alpha$& 0.0130& 0.0116& 0.0100 & 0.0064 \\
\bottomrule
\end{tabular}
}
\end{minipage}
\begin{minipage}{0.48\linewidth}
\captionof{table}{KL divergence obtained by SVRG with varying learning rate on DS1. We select $\rho=0.001$ in our implementation.}
\label{tab:kl-svrg}
\resizebox{\linewidth}{!}{
\begin{tabular}{ccccc}
\toprule
learning rate $\rho$ & 0.0001 & 0.0003 & 0.001 & 0.003 \\
\midrule
SVRG  & 0.0093 & 0.0094 & 0.0088 & 0.0118 \\
\bottomrule
\end{tabular}
}
\end{minipage}



\end{document}